\documentclass[twocolumn,times]{aastex631}
\usepackage[shortcuts]{extdash}
\usepackage[T1]{fontenc}
\usepackage{xcolor}

\begin{document}

\title{Joint Constraints on Exoplanetary Orbits from Gaia DR3 and Doppler Data}

\author[0000-0002-4265-047X]{Joshua N.\ Winn}
\affiliation{Department of Astrophysical Sciences, Princeton University, Princeton, NJ 08544, USA}

\begin{abstract}

The third Gaia data release includes a catalog of exoplanets and
exoplanet candidates identified via the star's astrometric motion.
This paper reports on tests for consistency
between the Gaia two-body orbital solutions and
precise Doppler velocities, for the stars currently
amenable to such a comparison.
For BD\=/17\,0063, HD\,81040, and HD\,132406,
the Gaia orbital solution and the Doppler data
were found to be consistent, and were fitted jointly to obtain
the best possible constraints on the planets' orbits
and masses.
Inconsistencies were found for 4 stars:
HD\,111232, probably due to additional
planets that were not included
in the astrometric model;
HD\,175167 and HR\,810, possibly due to
inaccurate treatment of non-Gaussian uncertainties in the
Gaia orbital solutions;
and HIP\,66074, for unknown reasons.
Consistency tests were also performed for HD\,114762,
which was reported in 1989 to have a brown dwarf
or exoplanet but has since been shown to be binary star.
The joint Gaia-Doppler analysis shows the secondary mass to be
$0.215\pm 0.013\,M_\odot$ and the orbital inclination
to be $3.63\pm 0.06$~degrees.

\end{abstract}

\section{Introduction} \label{sec:intro}

The astrometric technique for exoplanet detection is based on
sensing a star's reflex motion projected onto the sky plane.
It is the oldest method for generating exoplanet candidates
\citep[see, e.g.,][]{Strand1943},
and has recently come back to prominence
thanks to data from the \citet{GaiaDR3Summary2022}.
As part of Data Release 3 (DR3), \citet{Holl+2022}
presented
a catalog of candidate exoplanets that includes 73 astrometric detections.

The astrometric method for planet
detection is important because it can reveal
all of a planet's orbital parameters and uniquely specify its mass \citep[see, e.g.,][]{Quirrenbach2010}.
However, despite the unprecedented precision of Gaia astrometry,
the current data are only sensitive to giant planets, and even then, the signal-to-noise
ratios are typically modest ($\sim$10).
With current Doppler spectrographs, giant planets can be
detected with much higher signal-to-noise ratios,
but the Doppler data
do not reveal the orbital inclination
and give only a lower limit on the planet's mass. Thus, the two techniques
are complementary.

The \citet{GaiaDR3TwoBody2022} highlighted 11 astrometrically detected
exoplanets for which high-precision
Doppler data are already available,
and compared the published Doppler-based values
of the orbital period and eccentricity with
the values determined by Gaia. The good matches
served to validate the methods that
were used to derive orbital parameters from the Gaia astrometry,
although the team also noted some discrepancies.
In keeping with the desire for the
initial DR3 publications to be based solely
on Gaia data insofar as possible, the Gaia team
did not perform joint fitting
of the astrometric and Doppler data.
Doing so would allow for a more thorough evaluation of the consistency
between the two datasets, and would also provide the best possible
constraints on the planets' masses and orbital parameters.
That was the purpose of the work described in this paper.

While analyzing the planet-hosting stars,
the opportunity was also taken to revisit HD\,114762\,b,
the erstwhile exoplanet candidate discovered 
by \citet{Latham+1989}.
Astrometric data from Gaia has already
shown that the orbit is viewed nearly face-on and
that the companion's mass exceeds the minimum mass of a star
\citep{Kiefer+2019, Holl+2022},
but the results of
joint Gaia-Doppler fitting
have not yet been reported.

This paper is organized as follows.
Section~\ref{sec:sample} introduces the sample of stars,
Section~\ref{sec:methods} explains the methods of analysis,
and Section~\ref{sec:results} describes the results.
The results are briefly summarized and discussed in Section~\ref{sec:discussion}.

\section{Sample Selection}
\label{sec:sample}

A straightforward approach to determining the orbital parameters
would be to fit a two-body model to the combination of
radial-velocity and astrometric time-series data.
However, the Gaia time-series astrometric
data have not yet been released.
Instead, DR3 provides a summary of the results of fitting a two-body
model to the astrometric time-series data. 
This information is sufficient for present purposes,
but only when
the star's radial-velocity variations and astrometric motion
are both dominated by the effect of a single planet.
When multiple planets make detectable contributions to
the star's motion, the results of a two-body fit
are difficult to interpret, at best, and meaningless, at worst.
Therefore, the list of stars to be analyzed
was restricted to those for which
which useful Doppler data are available
and only a single giant planet has been
reported.

This narrowed down the list to 7 stars (in addition
to the special case of HD\,114762, described above).
For one of those stars, HIP\,66074, the
Gaia astrometry triggered the initial detection of the planet,
although a small amount of precise radial-velocity data
had previously been obtained
as part of a Doppler survey.
For the other 6 stars,
BD\=/17\,0063,
HD\,81040,
HD\,132406,
HD\,111232,
HR\,810,
and HD\,175167,
the planet was initially discovered via the Doppler method.
The stars that needed to be rejected on account of
having multiple known companions were
HD\,142 \citep{Wittenmyer+2012},
HD\,164604 \citep{Arriagada+2010},
and GJ\,876 \citep{Marcy+2001}.
Another star, HIP\,28193,
was excluded because the available
radial-velocity data are sparse
and affected by stellar activity \citep{Holl+2022}.

Table~\ref{tbl:rv} gives a summary of the Doppler data
gathered from the literature.
The results of the two-body fits to the Gaia data
were obtained from the 
{\fontfamily{lmtt}\selectfont
gaiadr3.nss\_two\_body\_orbits}
table at the Gaia archive.\footnote{\url{https://gea.esac.esa.int/archive/}}
The
{\fontfamily{lmtt}\selectfont
gaiadr3.astrophysical\_parameters} table \citep{Creevey+2022} 
was also consulted to obtain
the stellar properties determined by
the General Stellar Parametrizer from Spectroscopy
({\fontfamily{lmtt}\selectfont gspspec})
and Photometry
({\fontfamily{lmtt}\selectfont gspphot}),
as well as the
stellar mass from the Final Luminosity Age Mass Estimator
(FLAME).

\begin{deluxetable*}{lccccccc}
\label{tbl:rv}
%\tablewidth{290pt}
\tabletypesize{\small}
\tablecaption{Sources of radial-velocity data.}
%\tablenum{2}

\tablehead{
\colhead{Star} & \colhead{Instrument} &
\colhead{\# of} & \colhead{Median} & \colhead{Timespan} & \colhead{Time range} & \colhead{Reference} \\[-0.1in]
\colhead{name} & \colhead{name} &
\colhead{RVs} & \colhead{unc.\ (m/s)}  & \colhead{(days)} & \colhead{(year/month)} & \colhead{}
} 
\startdata
BD\=/17\,0063 & HARPS    & 26 & 1.5 & 1760 & 2003/10--2008/08 & \citet{Moutou+2009}\\
HD\,81040 & ELODIE       & 23 & 12 & 1210 & 2002/02--2005/05 & \citet{Sozzetti+2006} \\
HD\,81040 & HIRES        & 3 & 12 & 254 & 1999/04--2000/01 & \citet{Sozzetti+2006} \\
HD\,132406 & ELODIE      & 17 & 11 & 748 & 2004/05--2006/06 & \citet{DaSilva+2007} \\
HD\,132406 & SOPHIE      & 4 & 4 & 150 & 2006/12--2007/05 & \citet{DaSilva+2007} \\
HIP\,66074 & HIRES       & 10 & 1.5 & 1399 & 2009/04--2013/02 & \citet{Butler+2017} \\
HR\,810    & UCLES       & 25 & 4.7 & 2458 & 1998/10--2005/07 & \citet{Butler+2017} \\
HR\,810    & CES         & 95 & 17 & 1977 & 1992/11--1998/04 & \citet{Kurster+2000} \\
HR\,810    & CORALIE     & 26 & 9 & 604 & 1998/07--2000/03 & \citet{Naef+2001} \\
HR\,810    & HARPS       & 47 & 2.0 & 717 & 2003/11--2005/10 & \citet{Trifonov+2020} \\
HD\,175167 & MIKE        & 13 & 4.2 & 1828 & 2004/07--2009/07 & \cite{Arriagada+2010} \\
HD\,111232 & MIKE        & 15 & 3.2 & 1326 & 1998/02--2001/10 & \cite{Minniti+2009} \\
HD\,111232 & CORALIE     & 38 & 5.5 & 1181 & 2000/03--2003/06 & \cite{Mayor+2004} \\
HD\,111232 & HARPS       & 58 & 2.0 & 4489 & 2004/02--2016/05 & \cite{Trifonov+2020} \\
HD\,114762 & HJS/Coud\'{e} & 86 & 34 & 591 & 1988/11--1990/07 & \cite{Cochran+1991} \\
HD\,114762 & Hamilton$^\star$ & 74 & 22 & 6900 & 1990/03--2009/02 & \cite{Kane+2011} \\
HD\,114762 & HIRES       & 24 & 1.7 & 2039 & 2013/12--2019/07 & \cite{Rosenthal+2021}
\enddata
% table was scrubbed on 3 Sep 2022
\tablecomments{For comparison, the
Gaia observations that are the basis of DR3
took place over about 1038 days, between
25 July 2014 and 28 May 2017 \citep{GaiaDR3Summary2022}.}
\tablenotetext{*}{The dewar
for the Hamilton spectrograph was changed 6 times over the timespan of the observations.
Each dewar change was regarded as a change of spectrograph, introducing another
offset and jitter parameter.}
\end{deluxetable*}

\section{Methods}
\label{sec:methods}

For each star, there were three stages in the analysis.
First, the Doppler data were fitted alone 
(Section~\ref{subsec:doppler}).
Second, the Gaia data were fitted alone,
or to be more precise, the tabulated
results of the two-body fit were
used to determine the posterior probability
distributions for the parameters that are independently constrained
by the Doppler data
(Section~\ref{subsec:gaia}).
Third, the Doppler and Gaia results were compared in detail
(Section~\ref{subsec:joint}).
If they were consistent, then a joint fit was
performed. The results for individual stars are described in Section~\ref{sec:results}.

\subsection{Doppler analysis}
\label{subsec:doppler}

The Doppler data are in the form of one or several time
series of radial-velocity measurements $v(t_i)$
and associated uncertainties $\sigma_i$, with each series
coming from a different telescope and spectrograph.
They were modeled with the radial-velocity equation,
\begin{equation}
    v(t) = K \left\{ \cos [f(t) + \omega] + e\cos\omega \right\} + \gamma,
\end{equation}
where $K$ is the radial-velocity semiamplitude, $e$ is the orbital
eccentricity, $f(t)$ is the true anomaly, 
$\omega$ is the argument of pericenter,
and $\gamma$ is the time-independent component of the
radial velocity data (either the actual radial velocity of the center of mass
or, more typically, an arbitrary radial velocity associated with the template spectrum from
which relative radial velocities were determined).
Given the time of a measurement, the true anomaly
was calculated from $t_{\rm p}$ and $e$ 
by iteratively solving Kepler's equation for the eccentric anomaly $E(t)$,
\begin{equation}
  \frac{2\pi}{P}(t-t_{\rm p}) = E(t) - e\sin E(t)
\end{equation}
and then calculating
\begin{equation}
    f(t) = 2\tan^{-1}\left[ \sqrt{\frac{1+e}{1-e}} \tan \frac{E(t)}{2} \right].
\end{equation}

The Doppler likelihood function ${\mathcal L}_{\rm v}$ was taken to be
\begin{equation}
\label{eq:doppler-likelihood}
    {\mathcal L}_{\rm v} = \prod_{i=1}^N \frac{1}{\sqrt{2\pi (\sigma_{v,i}^2 + \sigma_0^2)}}
    \exp\!\left[-
    \frac{ (v_i - v_{i,\,{\rm calc}})^2 }{ 2(\sigma_{v,i}^2 + \sigma_0^2) }
    \right],
\end{equation}
where $i$ runs over all the data points,
$v_{i,{\rm calc}}$ is the calculated velocity
based on a given choice of model parameters, 
$v_i$ is the measured radial velocity,
$\sigma_{v,i}$ is the formal uncertainty,
and $\sigma_0$ is the `velocity jitter,' a constant meant to account for unmodeled
systematic errors.

Posterior sampling was performed with the
Monte Carlo Markov Chain algorithm of \citet{GoodmanWeare2010} as implemented by the code {\tt emcee} \citep{ForemanMackey+2013}.\footnote{Here and
elsewhere, the number of `walkers' was 32 and the number of `links' was chosen
to be 500{,}000, which was always more than 50 times the integrated autocorrelation length
for each parameter.}
The `stepping parameters' in the chain were
\begin{equation}
    \{ K, \sqrt{e}\cos\omega, \sqrt{e}\sin\omega, P, t_{\rm p} \}
\end{equation}
along with the nuisance parameters
$\gamma$ and $\sigma_0$ specific to the data from each spectrograph.
Thus, for a case in which all the data were from a single
spectrograph, the total number of parameters was 7. Each
additional spectrograph increased the number of parameters by 2.
Uniform priors were adopted for each stepping parameter.

\subsection{Gaia analysis}
\label{subsec:gaia}

The traditional or `Campbell' orbital elements are
\begin{equation}
\{a, e, I, \omega, \Omega, P, t_{\rm p}\},
\end{equation}
where $a$ is the semimajor axis,
$I$ is the inclination, and
$\Omega$ is the longitude of the ascending node.
The Gaia DR3 results are expressed in a different basis,
\begin{equation}
\label{eq:gaia-params}
\{A, B, F, G, e, P, t_{\rm p}\},
\end{equation}
where 
%$\varpi$ is the parallax and
$A$, $B$, $F$, and $G$ are the angular Thiele-Innes coefficients,
defined as
\begin{eqnarray}
A &=&  a_0\left( \cos\omega\cos\Omega - \sin\omega\sin\Omega\cos I\right), \\
B &=&  a_0\left( \cos\omega\sin\Omega + \sin\omega\cos\Omega\cos I\right), \\
F &=& -a_0\left( \sin\omega\cos\Omega + \cos\omega\sin\Omega\cos I\right), {\rm and} \\
G &=& -a_0\left( \sin\omega\sin\Omega - \cos\omega\cos\Omega\cos I\right).
\end{eqnarray}
In these definitions, $a_0$ is the semimajor axis of the
observed orbit
converted into angular units by multiplying by the parallax, $\varpi$.
For a star with a dark companion, such as a planet,
the observed orbit is the star's orbit. When the light from the
companion is not negligible, the observed orbit is that of the
`photocenter' (the apparent position of the unresolved
combination of light from both bodies), a point discussed further
in Section~\ref{subsec:joint}.

The Gaia team determined
the orbital elements by fitting a model to the Gaia time-series
astrometric data that
also included parameters for the
position, proper motion, and parallax.
The
{\fontfamily{lmtt}\selectfont
gaiadr3.nss\_two\_body\_orbits}
table
contains a list of best-fit parameters
and a correlation matrix, which was converted
into a covariance matrix $\mathbf{C}$ using the 
{\fontfamily{lmtt}\selectfont nsstools}
code\footnote{\url{https://www.cosmos.esa.int/web/gaia/dr3-nss-tools}}
\citep{Halbwachs+2022}.
The Gaia likelihood function ${\mathcal L}_{\rm g}$ was taken to be
\begin{equation}
    {\mathcal L}_{\rm g} = \frac{1}{\sqrt{(2\pi)^8 |{\rm det}\,\mathbf{C}}|} \exp\left[ -\frac{1}{2}\left( \mathbf{\Theta}^{\rm T} \mathbf{C}^{-1} \mathbf{\Theta} \right)^2 \right],
\end{equation}
where $\mathbf{\Theta}$ is the `Gaia deviation vector,' an 8-element column vector composed of differences
between the Gaia-measured values and the calculated values
for the 7 parameters given in Eqn.~(\ref{eq:gaia-params}) and the parallax.

The Gaia likelihood was used to produce samples from
the posterior probability density for the parameters
\begin{equation}
\{ a_0, e, \cos I, \omega, \Omega, P, t_{\rm p}, \varpi \}.
\end{equation}
Uniform priors were employed for these parameters,
and the
{\fontfamily{lmtt}\selectfont emcee}
code was used for sampling.
At each step in the chain, these parameters were used to compute
$A$, $B$, $F$, and $G$, thereby allowing $\mathbf{\Theta}$ to be constructed
and the likelihood to be evaluated.

\clearpage
\subsection{Joint analysis}
\label{subsec:joint}

At this stage, a comparison was made between the Doppler and Gaia-based results
for the set of parameters they have in common:
\begin{equation}
\{ e, \omega, P, t_{\rm p} \}.
\end{equation}
For the $t_{\rm p}$ parameter, the Gaia convention was followed, in which $t_{\rm p}$
refers to the orbit that was underway at epoch 2016.0
(JD~2{,}457{,}389.0), and takes values between $-P/2$ and $+P/2$.
In addition, the Doppler results for $K$
were compared to the radial-velocity
semiamplitude implied by the Gaia parameters (assuming the light
from the companion is negligible),
\begin{equation}
    K = \frac{2\pi}{P}\,\frac{(a_0/\varpi)\sqrt{1-\cos^2\!I}}{\sqrt{1-e^2}}.
\end{equation}
If the parameters were inconsistent, an attempt was made to
ascertain the reason, as described in the next section on a case-by-case basis.
If they were consistent, then a joint fit was performed.

The model parameters for the joint fit were
\begin{equation}
\{ M, m, \sqrt{e}\cos\omega, \sqrt{e}\sin\omega, \cos I, \Omega, P, t_{\rm p}, \varpi, \varepsilon \}
\end{equation}
along with the nuisance parameters $\gamma$ and $\sigma_0$ for each spectrograph.
A Gaussian prior was placed on the primary mass $M$ based on the
value from the
{\fontfamily{lmtt}\selectfont
gaiadr3.astrophysical\_parameters}
table,
or from the literature, as described below.
The secondary mass $m$ was subject to a uniform prior, as were
all the other parameters.
The $\varepsilon$ parameter is the flux ratio between the companion and the primary star.
The flux ratio is relevant
because, as noted earlier,
Gaia measures the motion of the center-of-light
of the star and any unresolved companions. Thus, strictly speaking, the 
reported orbital parameters pertain to the `photocenter' and not the
star. For a given choice of model parameters, $a_0$ was computed 
using the equation
\begin{equation}
    \frac{a_0}{\varpi} = [G(M+m)]^{1/3} \left( \frac{P}{2\pi} \right)^{\!\!2/3}
    \left( \frac{m}{M+m} - \frac{\varepsilon}{1+\varepsilon} \right),
\end{equation}
which is based on Kepler's third law and the assumption that the photocenter
is the flux-weighted mean position of the two bodies.
Once $a_0$ is determined, the Thiele-Innes coefficients can be computed, the
Gaia deviation vector $\mathbf{\Theta}$ can be constructed, and the likelihood
$\mathcal{L}_{\rm g}$ can be evaluated.

Although the correction for $\varepsilon$ is important for binary stars,
we expect it to be negligible ($\ll$\,$10^{-4}$) for planets and brown dwarfs.
For example, according to Table 5 of the
compilation of stellar properties by
\cite{PecautMamajek2013} (as updated on the 
website of
E.~Mamajek\footnote{\url{https://www.pas.rochester.edu/~emamajek/EEM_dwarf_UBVIJHK_colors_Teff.txt}}),
a solar-mass star with an 0.08\,$M_\odot$ companion
would have a Gaia-band flux ratio of 
about $4\times 10^{-5}$.
This allowed the $\varepsilon$ parameter to serve as another
consistency check: if the companion is a planet or brown dwarf
and the Gaia and Doppler data are
consistent, then the credible interval for 
$\varepsilon$ should encompass values smaller than $10^{-4}$.
Because of this firm expectation, whenever the data were found to be consistent
with a completely dark companion, the joint fit was repeated
with the additional constraint $\varepsilon = 0$. The intention
was to avoid biasing the results by allowing
unrealistically large flux ratios.

\section{Results}
\label{sec:results}

\subsection{BD\=/17\,0063}

BD\=/17\,0063 (HIP\,2247) is a K dwarf at a distance of 35 parsecs
with a Gaia optical apparent magnitude of $G=9.2$.
\cite{Moutou+2009}
discovered a giant planet around this star.
They determined the star's
mass to be
$0.74\pm 0.04\,M_\odot$ (the value adopted here),
and reported the
planet's minimum mass, period, and eccentricity
to be 5.1 Jupiter masses, 655.6~days, and 0.54, respectively.
Judging from the literature,
the planet has not received much attention since
its discovery.

Although the available radial velocities span three planetary orbits,
most of the data were obtained during
a single orbit. The Doppler-only analysis
showed that the spectroscopic orbit is well constrained,
with the strongest covariances seen between $P$ and $t_{\rm p}$, and
between $\sqrt{e}\cos\omega$ and $\sqrt{e}\sin\omega$.

The Doppler data and Gaia two-body orbital solution
for this system are harmonious.
The Doppler-only results for $K$, $P$, $t_{\rm p}$, $e$, and $\omega$
all agree to within 1.5-$\sigma$ with the Gaia-only
results. The Doppler-only results are more precise.
For example, the $K$ value is $172.5\pm 1.6$~m/s
based only on radial velocities, and $147^{+45}_{-24}$~m/s
based on the Gaia orbital solution.

\begin{figure*}
\begin{center}
\includegraphics[width=0.9\textwidth]{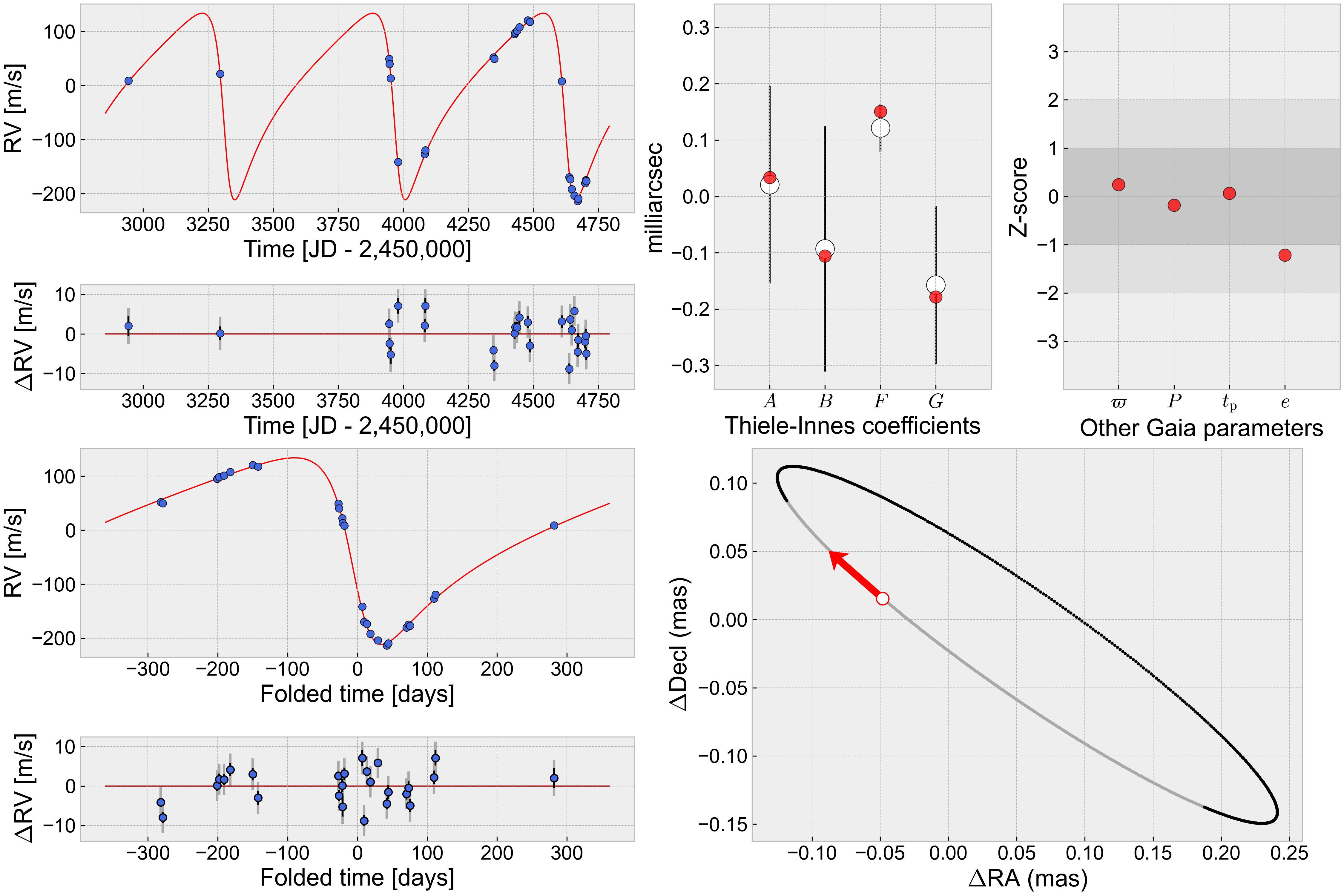}
\end{center}
\caption{\label{fig:BD-17_0063}
{\bf BD-17\,0063}:
Results of the joint Gaia-Doppler fit, assuming the companion is dark.
{\bf Left:} Doppler data as a function
of time and folded time, with residuals.
Red curves show the maximum {\it a posteriori}
probability (MAP) model.
{\bf Right:} The two top panels
compare the parameters of the joint model with those
from Gaia alone.
The first panel shows the Thiele-Innes coefficients;
open symbols are the Gaia-only values
and red points are from the MAP model.
The second panel shows the $Z$-score (`number of sigma')
between the Gaia-only and best-fit parameters.
The lower right panel shows the orbital geometry according
to the MAP model, with black representing the portion
on the `near side' of the sky plane and gray representing
the `far side'. The origin is the location of the center of mass,
the white circle marks the pericenter position,
and the red arrow conveys the direction of motion.
}
\end{figure*}

The joint fit was successful, providing a good match
to all of the radial velocities and the parameters of the Gaia two-body orbital
solution. The flux ratio was found to be compatible with zero,
with an upper bound of 0.18\% with 95\% confidence --- another sign of
concordance between the datasets. For this reason, the joint
fit was repeated with the constraint $\varepsilon=0$.
The quality of the fit can be assessed
in Figure~\ref{fig:BD-17_0063}, which shows the radial-velocity data
as a function of both time and phase (left), the level of
agreement with the Gaia orbital solution (upper right), and the orbital
geometry (lower right). For this system and the others,
Table~\ref{tbl:results} gives the results of the joint fit,
and
the Appendix contains
a `corner plot' showing the
two-dimensional posterior probability distributions for each pair
of parameters, marginalized over all the others.
In the joint fit, the orbit
is inclined by $81\pm 4$~degrees, the
planet's mass is $5.16\pm 0.20$ Jupiter masses, and the eccentricity is $e=0.5439\pm 0.0052$.

\begin{deluxetable*}{lccccc}
\label{tbl:results}
\tablecaption{Results of jointly fitting the Gaia and Doppler data.}
\tablehead{
\colhead{Parameter} &
\colhead{BD\=/17\,0063} & % scrubbed 04 Sep 2022
\colhead{HD\,81040} & % scrubbed 04 Sep 2022
\colhead{HD\,132406} & % scrubbed 04 Sep 2022
\colhead{HIP\,66074\tablenotemark{a}} & % scrubbed 04 Sep 2022
\colhead{HD\,175167\tablenotemark{b}} % scrubbed 04 Sep 2022
} 
\startdata
$M$ [$M_\odot$]     & $0.741\pm 0.040$          & $0.962\pm 0.040$               &
                      $0.973\pm 0.040$          & $0.671\pm 0.050$               & $1.090\pm 0.040$ \\
$m$ [$M_{\rm Jup}$] & $5.16\pm 0.20$            & $7.53\pm 0.032$                &
                      $5.94^{+0.65}_{-0.59}$    & $0.445^{+0.055}_{-0.046}$      & $14.8^{+1.8}_{-1.6}$ \\
$P$ [days]          & $655.57 \pm 0.58$         & $1004.7\pm 3.0$                &
                      $908\pm 16$               & $300.3\pm 1.3$                 & $1175\pm 25$  \\            
$t_{\rm p}$ [days]  & $-138.5\pm 2.8$           & $126\pm 17$                    &
                      $-225\pm 69$              & $52\pm 14$                     & $-194\pm 73$ \\
$e$                 & $0.5439\pm 0.0052$        & $0.527^{+0.034}_{-0.037}$      &
                      $0.250^{+0.063}_{-0.075}$ & $0.418\pm 0.067$               & $0.510\pm 0.035$ \\
$\cos I$            & $0.149\pm 0.069$          & $-0.366^{+0.077}_{-0.071}$     &
                      $-0.64^{+0.11}_{-0.09}$   & $-0.030^{+0.077}_{-0.081}$     & $0.814\pm 0.023$ \\
$\omega$ [rad]      & $1.987\pm 0.031$          & $1.275\pm 0.075$               &
                      $4.18\pm 0.22$            & $4.56\pm 0.018$                & $5.75^{+0.15}_{-0.18}$ \\
$\Omega$ [rad]      & $2.20\pm 0.13$            & $0.335^{+0.087}_{-0.083}$      &
                      $1.29\pm 0.25$            & $3.848^{+0.083}_{-0.089}$      & $1.15^{+0.14}_{-0.11}$ \\
$\varpi$ [mas]      & $28.99\pm 0.020$          & $29.01\pm 0.024$               &
                      $14.195\pm 0.016$         & $28.20\pm 0.011$               & $14.11\pm 0.019$ \\
$\varepsilon$ (\%)  & $\equiv 0$                & $\equiv 0$                     &
                      $\equiv 0$                & $1.00\pm 0.12$                 & $\equiv 0$                     
\enddata
%\tablerefs{reference list}
\tablecomments{$M$ is the primary mass; $m$ is the secondary mass, $P$ is the orbital period, $t_{\rm p}$ is the
Julian date of pericenter minus 2{,}457{,}389.0, $e$ is the eccentricity,
$\cos I$ is the cosine of the inclination,
$\omega$ is the argument of pericenter,
$\Omega$ is the longitude of the ascending node,
$\varpi$ is the parallax, and
$\varepsilon$ is the flux ratio.}
\tablenotetext{a}{Results for HIP\,66074 should be
interpreted cautiously because of the unrealistically high flux ratio.}
\tablenotetext{b}{Results for HD\,175167 should be
interpreted cautiously because of the tension between the Doppler-only
and Gaia-only orbital parameters.}

\end{deluxetable*}

\subsection{HD\,81040}

HD\,81040 (HIP 46076) is a G star at 34~pc with $G=7.6$.
According to the
Gaia team's Final Luminosity Age Mass Estimator
(FLAME), the star's mass
is $0.962\pm 0.040$\,$M_\odot$.
\citet{Sozzetti+2006} discovered a giant planet around this star with
the Doppler technique. 
They estimated the star's age to be
0.8~Gyr based on its chromospheric activity level and the
detection of lithium in its spectrum. \citet{Li+2021}
used a different activity/age calibration to arrive at
an age of $1.8\pm 0.3$~Gyr, which is also
consistent with the measured rotation period of
15-16 days. \citet{Li+2021} also spearheaded
the effort to combine Doppler and Gaia data for this star.
Although they did not have
the Gaia two-body solution at their disposal,
they constrained the three-dimensional orbital configuration 
by combining the radial-velocity data with a measurement of the
star's secular acceleration, which was in turn based on the
difference between the Hipparcos and Gaia proper motions.

The radial-velocity data span 3 planetary orbits 
but because of spotty time coverage, the parameters of the
Doppler-only fit were not constrained as well as they were
for BD\=/17\,0063.
In particular, the posterior distribution for the period is bimodal,
with a peak at the favored period
at about 1000~days and a secondary peak at about 1100 days
containing about 1/3 of the total probability.

Of the two periods, the 1000-day period agrees better
with
the period of the Gaia astrometric orbit ($833\pm 110$~days).
The two periods are about 1.5-$\sigma$ apart.
Likewise, the Doppler-only and Gaia-only values of the eccentricity agree to
within 1.5-$\sigma$, with $e=0.530^{+0.048}_{-0.073}$
from the Doppler
data and $0.35\pm 0.15$ from the Gaia orbital solution.
The time of pericenter, argument of pericenter,
and $K$ values agree to within 1-$\sigma$.

\begin{figure*}
\begin{center}
\includegraphics[width=0.9\textwidth]{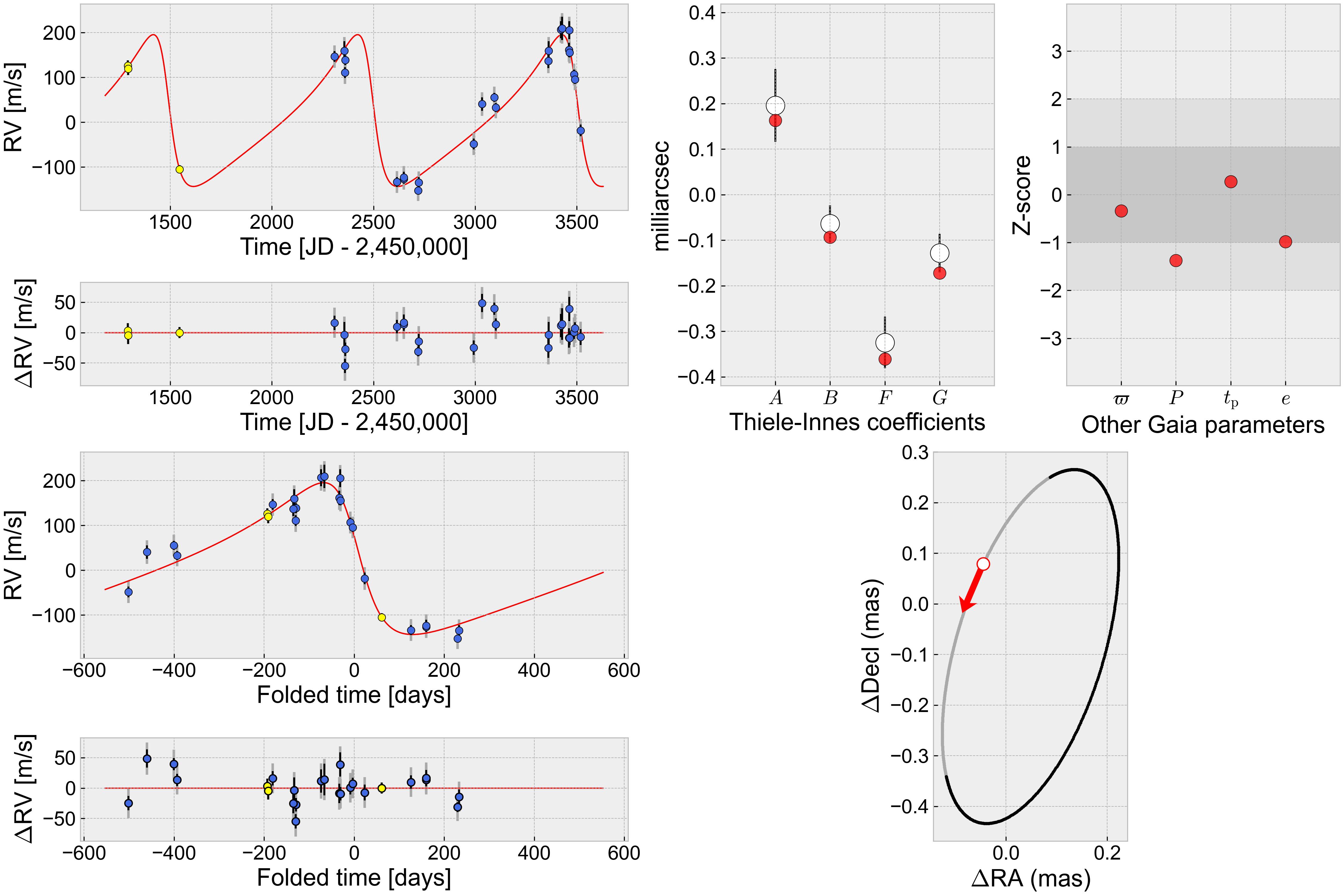}
\end{center}
\caption{\label{fig:HD81040}
{\bf HD~81040}:
Results of the joint Gaia-Doppler fit, assuming the companion is dark.
Same format as Figure~\ref{fig:BD-17_0063}.
In the radial-velocity plots, the filled symbols
are from ELODIE and the open symbols are from HIRES.
}
\end{figure*}

Thus, a joint fit seemed warranted.
The best-fit model
was able
to reproduce all of the relevant characteristics of the Doppler data
and the Gaia orbital solution.
The flux ratio was bounded to be below
0.098\% with 95\% confidence, another mark of 
consistency.
For the final results, the joint fit was repeated
under the constraint $\varepsilon=0$.
Figure~\ref{fig:HD81040} shows the fit
in the same format as in Figure~\ref{fig:BD-17_0063}, and the corner plot can be found in the Appendix.
The orbital period is $1004.7\pm 3.0$~days, 
the planet's mass is $7.53\pm 0.32$ Jupiter masses,
and the orbital eccentricity is
$0.527^{+0.034}_{-0.037}$.
The inclination is $111.4^{+4.4}_{-4.7}$ degrees.

The results of the joint fit are also compatible with those of 
\citet{Li+2021}, who found a planet mass of $7.24^{+1.0}_{-0.37}$ Jupiter
masses and an eccentricity of $0.525\pm 0.025$.
Their results for the inclination were subject to a two-way discrete degeneracy,
with $I = 73^{+12}_{-16}$ or $107^{+16}_{-12}$ degrees,
the latter of which agrees with the results presented here.

\subsection{HD\,132406} 
\label{subsec:HD132406}

\cite{DaSilva+2007} reported a giant planet orbiting HD\,132406
(BD$+$53\,1752, HIP 73146),
a G star at a distance of 70\,pc with $G=8.3$.
They selected this star for their survey on account of its
relatively high metallicity ([Fe/H]~$=+0.18\pm 0.05$).
The team was trying to exploit the strong association
between metallicity and giant-planet occurrence in order
to find as many giant planets as possible.
They reported $m\sin I = 5.61\,M_{\rm Jup}$, $e=0.34$,
and $P=974$ days. The literature has remained quiet about
this star since 2007. For this
study, a stellar mass of $0.973\pm 0.040$\,$M_\odot$ was adopted,
based on the FLAME results.

This was a case in which 
neither the Doppler data nor the Gaia data
were very constraining by themselves.
The Doppler data
extend over only one orbital cycle
and the velocity extrema are poorly covered.
The Doppler-only analysis
gave $K=117^{+179}_{-22}$ m/s.
Likewise, Gaia
detected the astrometric
orbit with a relatively
modest statistical
significance ($a_0=0.172^{+0.058}_{-0.034}$~mas),
leading to a predicted $K$ value
of $131^{+80}_{-53}$ m/s.

The Doppler and Gaia results for the $K$ parameter and all of the other
parameters they have in common were found to be in agreement,
justifying
a joint fit.
This pinned down the $K$ value to
$101\pm 10$~m/s.
However, the Gaia data
cannot strongly exclude a
nearly face-on orbit; in the joint fit,
the cosine
of the inclination is $-0.81\pm 0.14$.
As a result, the posterior probability distribution
for the companion mass has a long tail extending to
high values. The marginalized result is 
$m=7.9^{+6.8}_{-1.6}$
Jupiter masses, extending into the brown-dwarf
regime. Higher companion masses are
associated with higher flux ratios,
following the approximate relation
$\varepsilon \approx 0.9\times 10^{-3}\,m/M_{\rm Jup}$. 
Essentially, the orbit can be nudged closer to face-on
if the companion mass is increased (to keep $m\sin I$
constant and maintain agreement with the Doppler data)
and the flux ratio is increased
(to preserve the size of the observed orbit
and maintain agreement with the Gaia data).

To prevent the results from being influenced by
statistically acceptable models with unrealistically high flux ratios,
the fit was repeated
under the constraint $\varepsilon = 0$.
This broke the $m/I$ degeneracy, leading to the results
$m=5.94^{+0.65}_{-0.59}\,M_{\rm Jup}$
and $\cos I = -0.636^{0.113}_{-0.086}$.
The orbit is mildly eccentric, with 
$e=0.250^{+0.063}_{-0.075}$.
The results are given in Table~\ref{tbl:results} and depicted
in Figure~\ref{fig:HD132406}.
A corner plot can be found in the Appendix.

\begin{figure*}
\begin{center}
\includegraphics[width=0.9\textwidth]{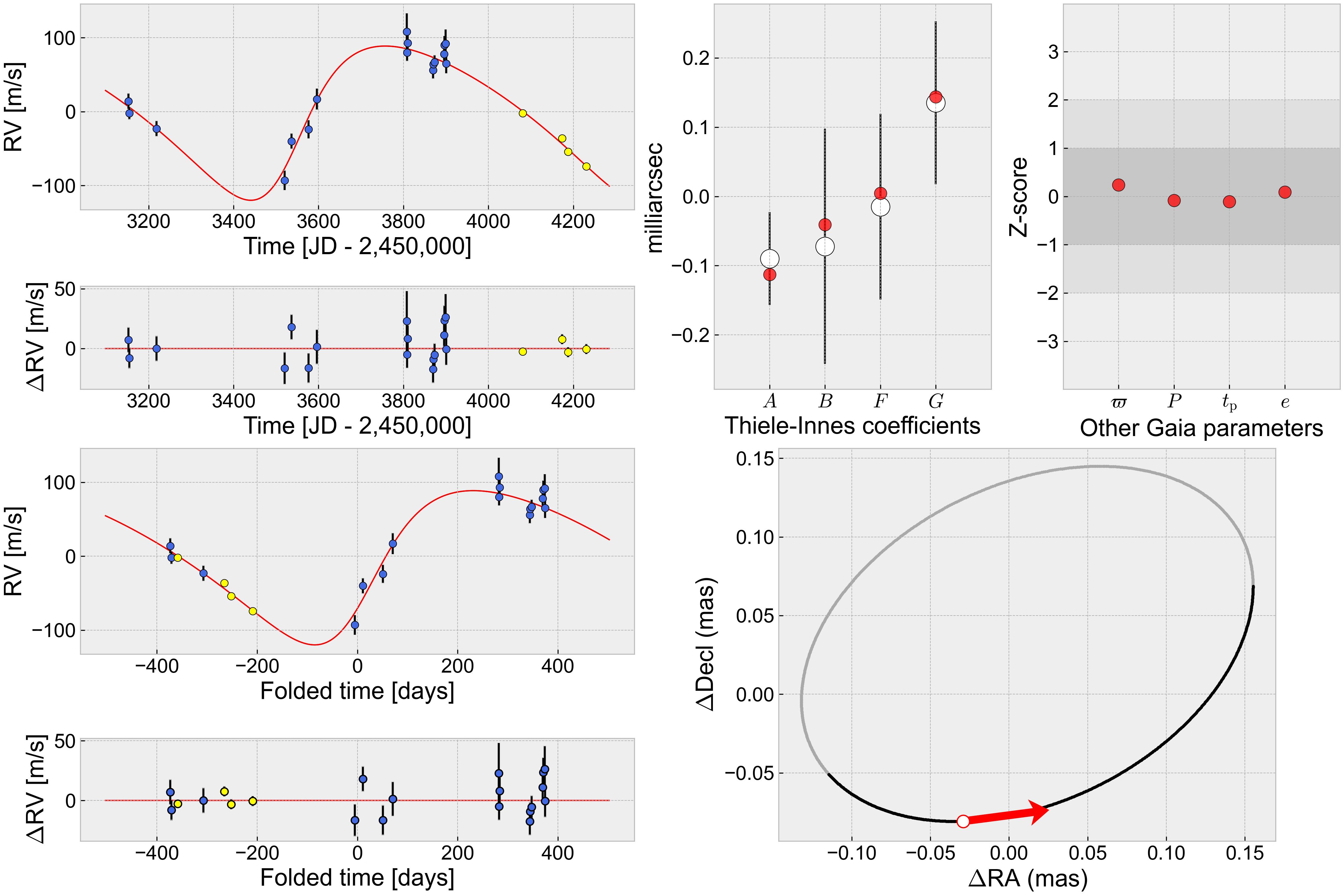}
\end{center}
\caption{\label{fig:HD132406}
{\bf HD~132406}:
Results of jointly fitting the Doppler data and Gaia two-body orbital solution,
assuming the companion is dark.
Same format as Figure~\ref{fig:BD-17_0063}.
In the radial-velocity plots, the filled symbols
are from ELODIE and the open symbols are from SOPHIE.
}
\end{figure*}

\subsection{HIP\,66074}

HIP\,66704 (BD$+$75\,510, GJ\,9452) is a K dwarf
at a distance of 35 parsecs with a Gaia optical magnitude of $G=9.7$.
The 
{\fontfamily{lmtt}\selectfont
gaiadr3.astrophysical\_parameters}
table does not provide a mass
estimate, but does provide a spectroscopy-based
effective temperature of $4161\pm 3$\,K
({\fontfamily{lmtt}\selectfont
teff\_gspspec}).
The absolute $G$ magnitude implied
by the apparent magnitude and parallax is $7.0$.
A mass of $0.67\pm 0.05\,M_\odot$ was
adopted for this star,
based on the 
effective temperature, the absolute magnitude, and the online
version of Table 5 of \cite{PecautMamajek2013}.

The star was monitored in the Lick-Carnegie Exoplanet Survey, although no planet
had been announced prior to Gaia DR3.
The 10 available radial
velocities were taken from the catalog of
\citet{Butler+2017}.\footnote{The catalog
includes an 11th data point based on a
spectrum with a much lower signal-to-noise ratio than the others, which
was not used here. The time stamp of the omitted point is 2455042.76395.}
The Gaia team reported that the star
shows astrometric motion consistent with the presence of a giant
planet with a mass
of $7.3\,M_{\rm Jup}$ and $P=297$~days \citep{GaiaDR3TwoBody2022,Holl+2022}.

For the Doppler-only analysis, a
good fit was found with $P=301\pm 5$~days,
$K=19^{+14}_{-3}$ m/s,
$t_{\rm p} = 62_{+32}^{-37}$~days,
$e=0.45\pm 0.22$, and $\omega=261^{+23}_{-30}$~degrees.
As is typical with meager Doppler datasets,
the joint posterior probability distribution
includes a long tail encompassing models
with high $K$ and high $e$ that produce
`velocity spikes' during
time ranges when no data were obtained.

According to
the Gaia two-body orbital solution,
$P=297.9\pm 2.7$~days, $e=0.40\pm 0.17$,
$t_{\rm p} = 58\pm 26$~days,
and $\omega = 265\pm 19$~degrees.
All of these parameters are within 1-$\sigma$
of the corresponding Doppler-only parameters.
However, there is a
serious problem with the radial-velocity semiamplitude.
The Gaia-only value is $K=297^{+82}_{-62}$~m/s,
15 times (4.4-$\sigma$) higher than the Doppler-only
value.

\begin{figure*}
\begin{center}
\includegraphics[width=0.9\textwidth]{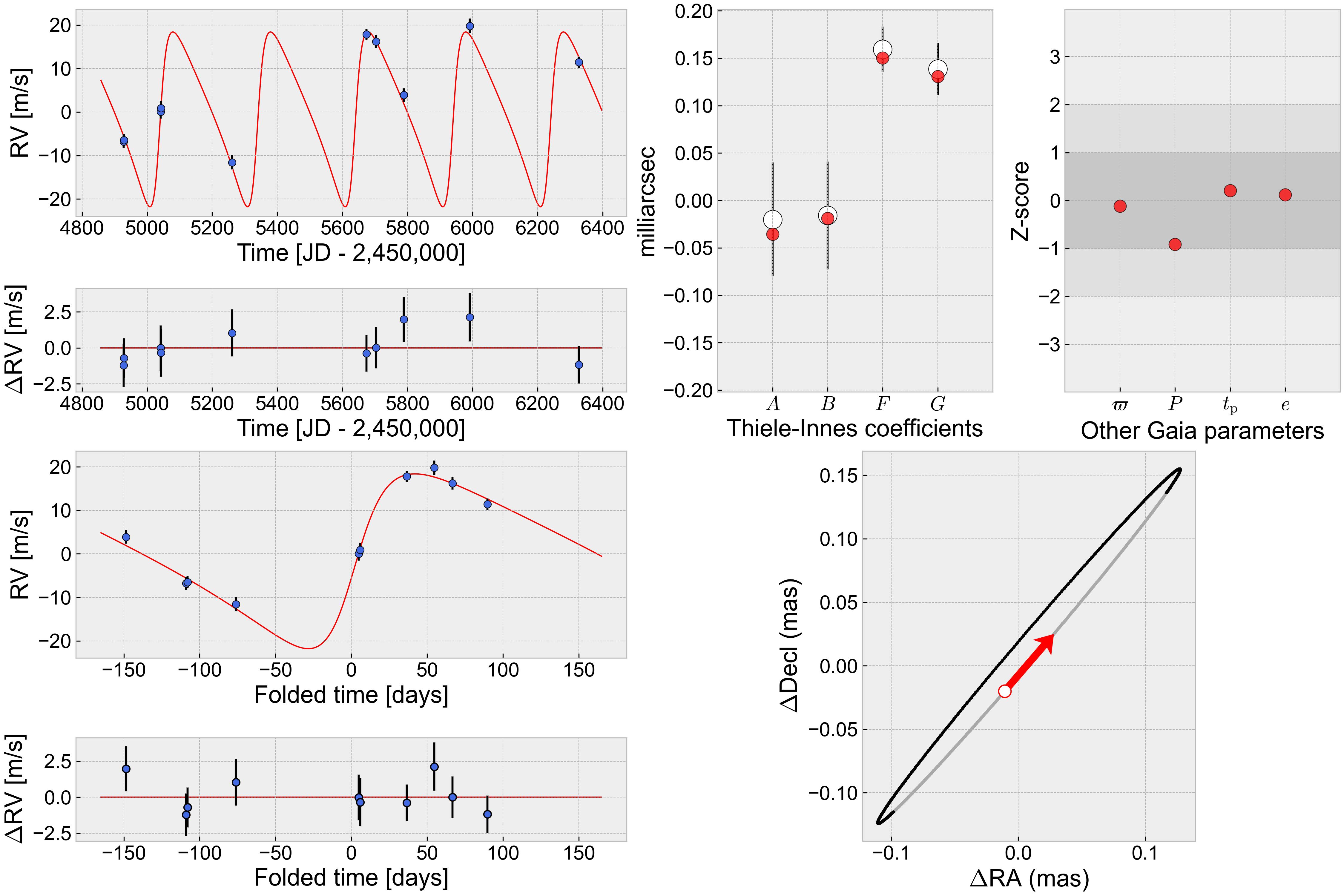}
\end{center}
\caption{\label{fig:HIP66074}
{\bf HIP\,66074}:
Results of jointly fitting the Doppler data and Gaia two-body orbital solution.
Same format as Figure~\ref{fig:BD-17_0063}.
}
\end{figure*}

Because of this problem, it is unclear whether a joint
fit is justified, although it was performed
anyways. In fact, an excellent
fit was achieved to all of the data (see Figure~\ref{fig:HIP66074}).
The sole peculiarity is that
the flux ratio is $0.0100\pm 0.0012$, i.e., incompatible with zero.
The effect of the nonzero flux ratio
is to cause the motion of the 
center of light --- the motion that is tracked
by Gaia --- to be reduced relative to the motion
of the primary star.
This allows the orbit of the primary
star to be wider and the orbital speed
to be slower, thereby reconciling the Gaia orbital solution with the
low $K$ value implied by the Doppler data.

A flux ratio on the order of $10^{-2}$ does not seem realistic, though,
given that the mass ratio was found to be $(6.3^{+0.7}_{-0.6}) \times 10^{-4}$.
No reasonable mass/luminosity relationship for brown
dwarfs or planets would
predict an optical flux ratio that is more than an order of magnitude higher
than the mass ratio.

Another way to state the problem is as follows.
The Doppler data require that $m\sin I = 0.435^{+0.038}_{-0.078}$
Jupiter masses, and the Gaia two-body solution involves a high inclination ($I=91\pm 5$~deg),
thereby cementing the companion's mass 
within the planetary-mass regime
and leading to a strong presumption that 
the flux ratio is $\ll 10^{-4}$.
However, the semimajor axis of the astrometric orbit measured by Gaia is too large
to be compatible with a dark planetary-mass object.
The tension is at the 10-$\sigma$ level.

The reason for the discrepancy is unclear.
Perhaps the system is home to more than just
a single companion and the results of
Doppler analysis, the Gaia two-body solution, or both,
were thrown off by the unmodeled
effects of additional bodies. Unfortunately,
there are not enough Doppler data, and not enough information in Gaia DR3,
to consider this hypothesis in detail.
Obtaining more Doppler data would help to clarify the
situation.

\subsection{HD\,175167} 

HD\,175167 (HIP\,93281) is a metal-rich G star ([Fe/H]$~=+0.19$)
located 71 parsecs away, with $G=7.8$ and a FLAME-based mass
of $1.09\pm 0.04\,M_\odot$.
\cite{Arriagada+2010} used the Doppler technique to
identify a giant planet
with $K=161$~m/s,
$P=1290$~days, and
$e=0.54$, which together give $m\sin I = 7.8$ Jupiter
masses. Only 13 velocities are available, and there are no other
accounts of precise Doppler observations in the literature.

The available Doppler data span only 1.4 orbital periods,
leading to a strong covariance between the uncertainties
of $P$ and $t_{\rm p}$.
The posterior probability distribution for the orbital period is
highly skewed. Marginalizing over $t_{\rm p}$ and all other parameters,
the Doppler-only estimate for the orbital period is $1283^{+14}_{-42}$ days.
The posterior for the orbital eccentricity is skewed, too,
giving $e=0.536^{+0.149}_{-0.035}$.
As was the case with
HIP\,66074, the Doppler-only fit admits the possibility of
high-$e$ solutions with huge velocity excursions when nobody was looking.

The Gaia DR3 orbital parameters are subject to unusually large
uncertainties, probably because the timespan of the Gaia observations (1038 days)
is shorter than the planet's orbital period.
According to the Gaia two-body orbital solution,
$P=899\pm 198$~days and $e=0.19\pm 0.12$.
The Gaia period is about 2-$\sigma$ lower than the Doppler-only period,
and the Gaia eccentricity is about 3-$\sigma$ lower than the Doppler-only
eccentricity, although such `number-of-sigma' comparisons are often misleading
when the uncertainty distributions are non-Gaussian.
The stated uncertainty in the time of periastron,
737 days, is comparable to the period itself.
Given that the uncertainties in the Gaia orbital parameters
are large and probably non-Gaussian,
this may be a case in which the tabulated best-fit values
and correlation matrix elements do not provide
enough information for an accurate treatment of the uncertainties.

With this caveat in mind, a joint fit was undertaken.
The model was able to provide a good
fit to the RV data and was also successful
in reproducing the measured Thiele-Innes
coefficients. The joint-fit eccentricity
is about 2.5-$\sigma$ higher
than the Gaia-only
value, a sign of unrelieved stress between
the Doppler data and the Gaia orbital solution.
The joint fit gives a companion mass
of $18.1^{+5.5}_{-2.9}$ Jupiter masses,
suggesting that it the companion might be better designated as a brown dwarf
than a giant planet. The flux ratio converged
on small values, with a 95\%-confidence upper bound of 1.5\%.

\begin{figure*}
\begin{center}
\includegraphics[width=0.9\textwidth]{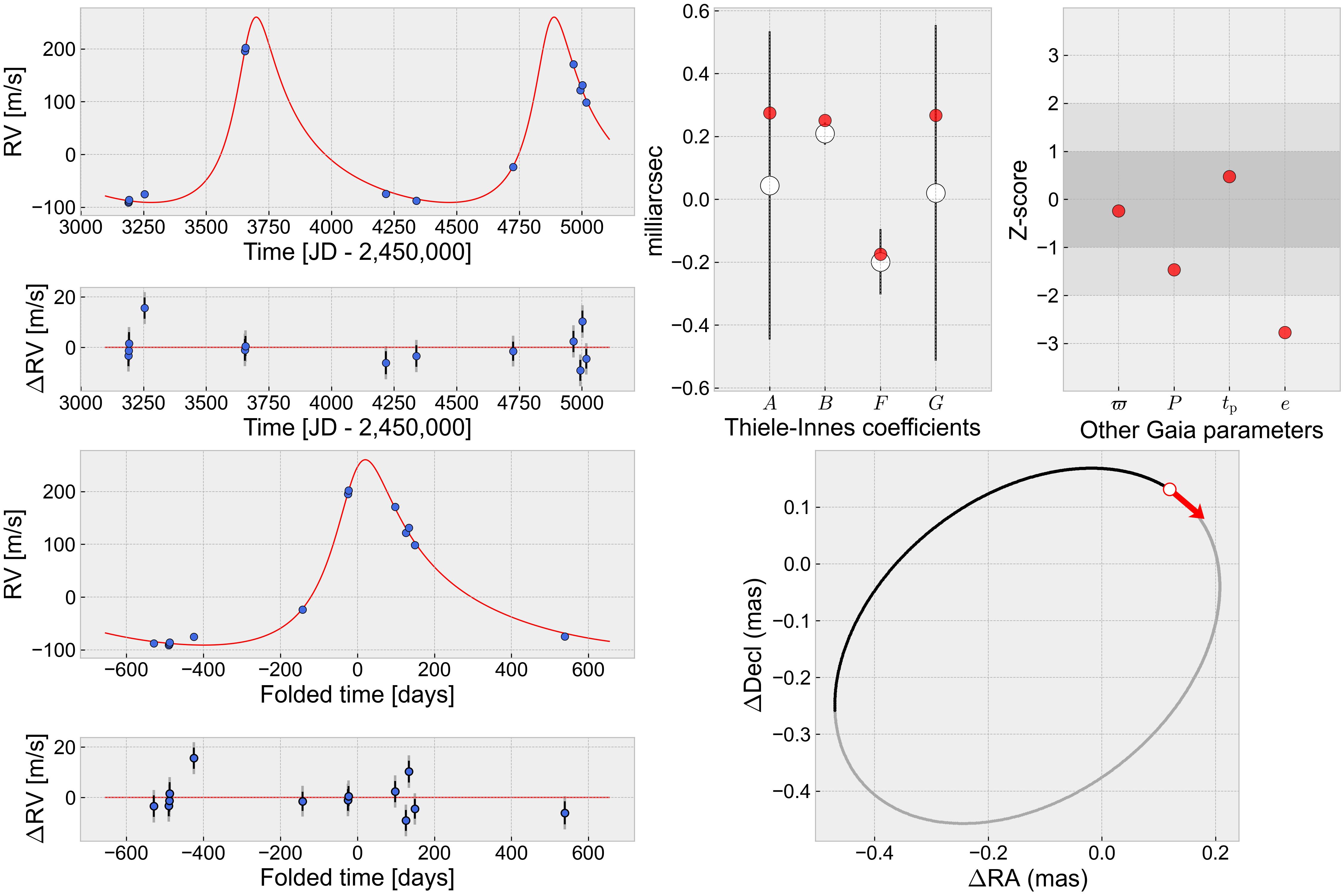}
\end{center}
\caption{\label{fig:HD175167}
{\bf HD\,175167}:
Results of jointly fitting the Doppler and astrometric data. Same format as Figure~\ref{fig:BD-17_0063}.
}
\end{figure*}

As was the case for HD\,132406, the allowed region
in parameter space includes a thin `branch' of solutions with very low inclinations, relatively
high companion masses, and unrealistically high flux ratios ($\sim$1\%).
To suppress this solution branch,
the joint fit was repeated under the constraint $\varepsilon=0$.
The companion's mass was thereby pinned
down to be $14.8^{+1.8}_{-1.6}$ Jupiter masses,
with a corresponding
orbital inclination of $35.5\pm 2.3$~degrees.
The results are given in Table~\ref{tbl:results}, Figure~\ref{fig:HD175167},
and the Appendix. They should be interpreted cautiously, because of
the large and non-Gaussian uncertainties in the Gaia orbital
solution.

\subsection{HR\,810}

HR\,810 (Iota Horologium, HD\,17051; HIP\,12653 GJ\,108) is a G0
star 17~pc away that shows X-ray and
Ca\,II emission characteristic of young stars.
\cite{Kurster+2000} reported
that the star is on the zero-age main
sequence, with an age between 30\,Myr and 2\,Gyr.
% JNW
Long-term monitoring of its ultraviolet and X-ray flux, and
chromospheric emission lines, has given
evidence for stellar activity cycles with durations of a few
years \cite[see, e.g.][]{Flores+2017,Sanz-Forcada+2019}.
% JNW
The star's mass is $1.077\pm 0.040$ according
to the Gaia DR3 FLAME code.
With a Gaia optical magnitude of $G=5.3$, 
HR\,810 is brighter than the other
stars analyzed in this work.

A giant planet orbiting HR\,810 with a minimum mass of 2.2~$M_{\rm Jup}$
and a period of 320 days
was discovered by \citet{Kurster+2000} using the Doppler
technique, and confirmed with additional data
from \citet{Butler+2001} and \citet{Naef+2001}.
Additional Doppler data are also available
in the user-friendly HARPS archive created by \citet{Trifonov+2020}.

\begin{figure*}
\begin{center}
\includegraphics[width=0.9\textwidth]{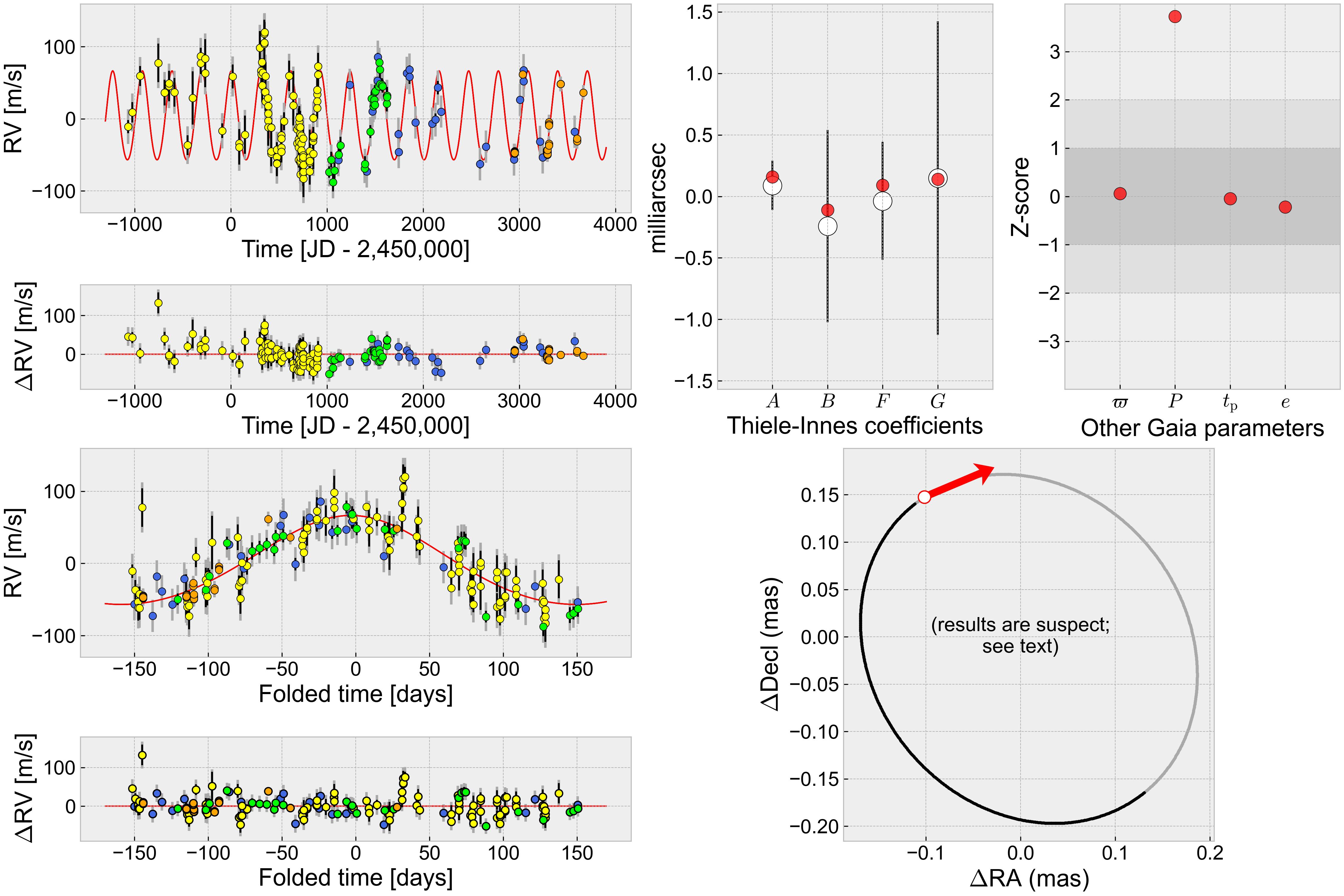}
\end{center}
\caption{\label{fig:HR810}
{\bf HR\,810}:
Results of jointly fitting the Doppler and astrometric data. Same format as Figure~\ref{fig:BD-17_0063}. The results should
be viewed skeptically; see the text for details.
}
\end{figure*}

According to the Doppler-only analysis,
the radial-velocity semiamplitude and orbital period
are well constrained, with $K=61.1\pm 2.5$~m/s
and $P=308.8\pm 0.6$~days. The orbit is nearly circular,
with $e=0.105^{+0.040}_{-0.046}$.
However, as is typical of young and chromospherically active stars,
the Doppler data appear to be affected by systematic errors
in excess of the formal measurement precision. The model
responds by enlarging the `jitter' parameters
as needed, but the accuracy of the results hinges on the assumption
implicit in Eqn.~(\ref{eq:doppler-likelihood})
that the errors are independently drawn from
a time-invariant Gaussian
distribution. In this case, the residuals are
correlated in time, as can be seen in
Figure~\ref{fig:HR810}.
In addition to the large scatter observed in the residuals,
there are hints of a 1400-day periodicity,
with maxima at time coordinates 200, 1600, and 3000.
These undulations could be due
to an additional planet or stellar activity.

For HR\,810, the 
tabulated `astrometric jitter' parameter (which plays
the same role in the astrometric fit as the `velocity
jitter' does in the Doppler fit) is 0.279~mas, about three
times larger than for any of the other stars
analyzed in this work. However, by itself, the relatively high jitter is
not necessarily a symptom of a problem with the Gaia two-body solution.
Jitter values between 0.2 to 0.5~mas
are typical for stars
as bright as HR\,810, which exceed the
nominal Gaia bright limit of $G= 5.7$ \citep[see Fig.~A.1 of][]{Lindegren+2021}.

More worrying is that the tabulated Gaia orbital elements for this star
are highly uncertain. All of the Thiele-Innes coefficients 
are compatible with zero to within the formal
uncertainties. This fact may be related to an issue that
was described by \citet{GaiaDR3TwoBody2022}:
when the eccentricity is low, the two-body fitting code
tends to overestimate the uncertainties in the Thiele-Innes coefficients.
This issue led \cite{Babusiaux+2022}
to warn that `the covariance matrix for very low eccentricity solutions may be problematic.'
Thus, this is probably another case in which knowledge of the tabulated best-fit values
and correlation matrix is insufficient for an accurate treatment of
the uncertainties.

The Gaia-only period is $333.0\pm 5.8$~days,
which is close enough to the Doppler-only period that it is
unlikely to be a coincidence. However, the two periods
disagree by 4-$\sigma$. The Gaia-only prediction for $K$
is $205^{+102}_{-31}$~m/s, another $4$-$\sigma$
discrepancy with the Doppler-only analysis.
The Gaia-only eccentricity is $0.14^{+0.15}_{-0.10}$,
which does agree with the Doppler-only analysis.

Given the reasons to be skeptical of both the
Doppler and the Gaia analyses, the quantitative
results of the joint fit must be taken with a grain
of salt. Indeed, in the joint fit, the compromise that was struck between the Doppler data
and the Gaia orbital solution seems unsatisfactory. The orbit
was found to be nearly face-on ($\cos I = 0.957^{+0.037}_{-0.069}$,
or $I=17\pm 10$~degrees), even though the Gaia-only
fit prefers an edge-on orbit ($\cos I = 0.050\pm 0.085$).
The joint solution exhibits the same strong degeneracy
between $m$, $I$, and $\varepsilon$ that was seen for HD\,132406
(Section~\ref{subsec:HD132406}).

Because of the patterned residuals in the Doppler data,
the large formal uncertainties in the Gaia two-body solution,
and the statistical disagreement
between the Doppler-only and Gaia-only analyses, the
quantitative results for the joint fit are 
not given in Table~\ref{tbl:results},
although they are depicted in Figure~\ref{fig:HR810}
and in the Appendix.

\subsection{HD\,111232} 
 
HD\,111232 (HIP\,62534) is a G star located 29 parsecs away with
a Gaia optical magnitude of $G=7.4$.
Its mass is $0.897\pm 0.04$\,$M_\odot$, based on the FLAME parameters tabulated in Gaia DR3.
\cite{Mayor+2004} discovered a giant planet around this star, as part
of a survey with the CORALIE spectrograph.
The star is relatively metal
deficient, with [Fe/H]~$=-0.36$.
Based on the low metallicity as well as a high space velocity
of 104.4~km/s, \cite{Mayor+2004} proposed that the star
belongs to the galaxy's `thick disk.'
Their best-fit Doppler orbit had $P=1118$ days, $e=0.19$,
and $m\sin I = 6.7$ Jupiter masses.

\begin{figure*}
\begin{center}
\includegraphics[width=0.9\textwidth]{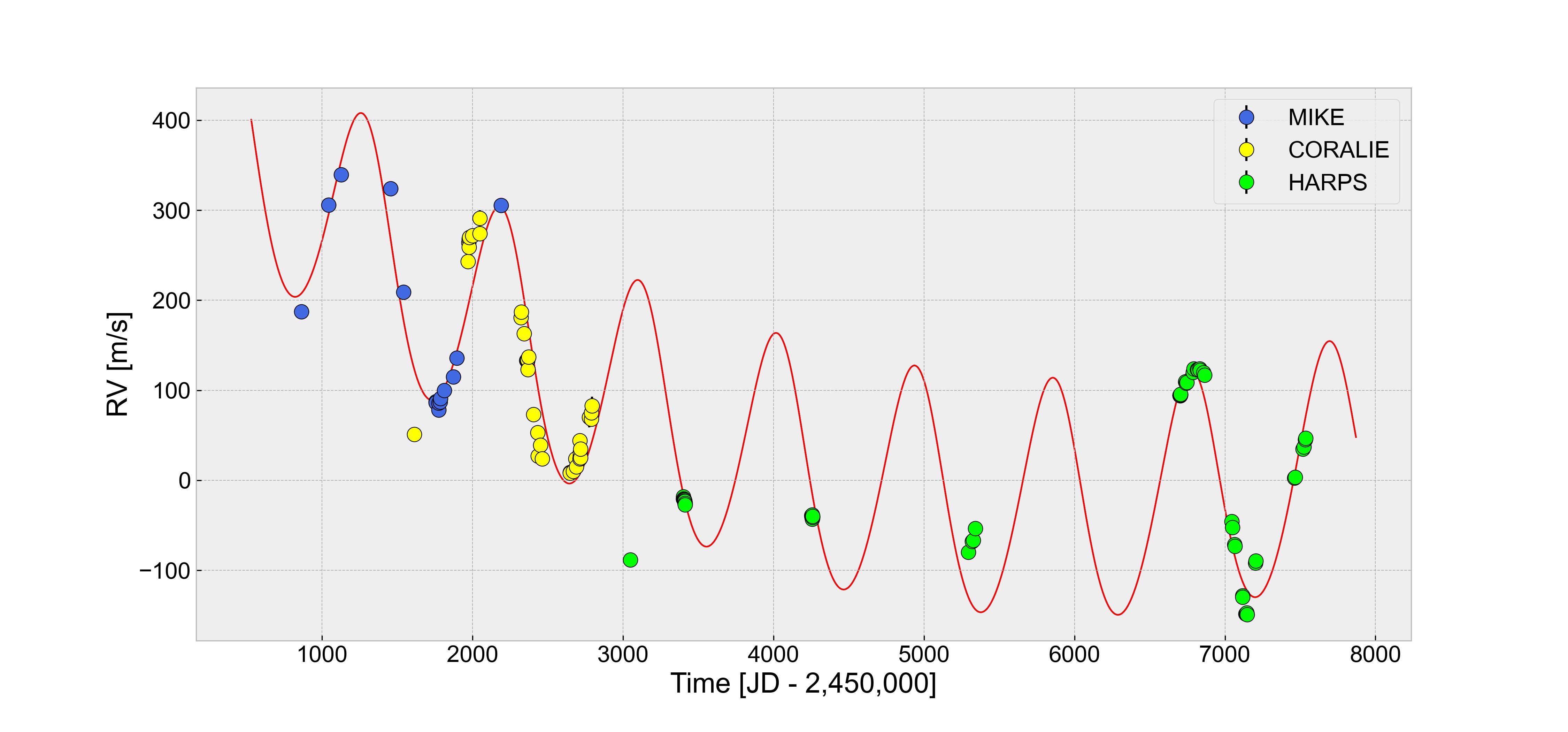}
\end{center}
\caption{\label{fig:HD111232}
{\bf HD\,111232}:
Doppler data and the best-fit model, including a single planet and an {\it ad hoc} quadratic function of time
representing the effects of other bodies.
}
\end{figure*}

The star was also observed by \citet{Minniti+2009}
with the MIKE spectrograph
as part of the Magellan Planet Search Program.
Their independent dataset
provided strong confirmatory evidence for the planet.
However, they did not fit the CORALIE and MIKE
data simultaneously. Had they done so, they would have
found that the combined dataset is
incompatible with a single-planet solution.
Additional data from the HARPS public
archive prepared by \cite{Trifonov+2020} confirms
this conclusion. Figure~\ref{fig:HD111232} shows all of the
radial-velocity data.

The red curve is the best-fitting
model involving a single planet and an {\it ad hoc}
quadratic function of time. The model provides
a decent qualitative description of most of the data,
but the residuals show clear patterns and major outliers.
In this model,
$P = 917$~days, which is closer
to the Gaia-derived value of $882\pm 30$~days 
than the originally reported periods of
1118~days \citep{Minniti+2009} and 
1143~days \citep{Mayor+2004}.
The model's eccentricity of $0.081$
is inconsistent with the Gaia-derived eccentricity
of $0.5\pm 0.1$.
% Note that I tried MCMC with radvel and it gave nonsense results.

Clearly, there is more to this system
than a star and a single giant planet. A joint fit was not
performed because the results of such a fit would be incoherent.
The star's additional
motion is obvious in the Doppler data and could also have
affected the astrometric measurements over the $\approx$1000-day timespan
of the Gaia observations.

\subsection{HD\,114762}

HD\,114762 (BD$+$18\,2700, HIP\,64426) is an early-G or late-F star
at a distance of 39 parsecs, with $G=7.1$.
There are discrepant reports of its effective temperature:
$5673\pm 44$\,K \citep{Kane+2011},
$5730_{-130}^{+37}$\,K ({\tt gspspec\_teff}),
$5837\pm 31$\,K \citep{Ghezzi+2010},
$5869 \pm 13$\,K \citep{Stassun+2017},
and $5935\pm 1$\,K ({\tt gspphot\_teff}).
Perhaps the differences are related to the star's
low metallicity [$-0.77 \pm 0.03$, per \citet{Kane+2011},
or $-0.66\pm 0.02$, per \citet{Sousa+2021}].
The Gaia DR3 FLAME mass, $1.047\pm 0.040$\,$M_\odot$,
is adopted here.

HD\,114762 played an important role in the history
of exoplanetary science. \cite{Latham+1989} found radial-velocity variations with 
a period of 84 days, an eccentricity of 0.3, and an implied $m\sin I$ of 11 Jupiter masses.
The authors wrote that the companion 
was `a good candidate to be a brown dwarf or even a giant planet.'
Either discovery would have been the first of its kind.

The prolonged debate over whether the companion was likely to be an exoplanet took some
interesting turns. Initially, it was unclear whether a giant planet could
have such a high mass, high eccentricity, and short period.
Of course, we now know of many giant planets with
high eccentricities and short periods, and there are many objects with masses of
10--20 Jupiter masses that are classified as exoplanets in the NASA Exoplanet Archive.

By the early 2000s, the occurrence of 
short-period giant planets was known to be strongly associated with
high metallicity \citep{Santos+2001,FischerValenti2005}.
This allowed another argument to be lodged against planethood for the companion of the
metal-poor star HD\,114762. 

In addition, there was the generic
problem common to all Doppler planets: the unknown inclination angle leaves
open the possibility that
the orbit is being viewed at low inclination, and $m$ is much larger
than $m\sin I$.
In an attempt to constrain the inclination, \citet{Cochran+1991} placed
an upper bound of 1 km/s on the star's projected rotation
velocity, which is anomalously low 
for stars of the same spectral type.
This suggested that the star's rotation axis has a low inclination and --- if the orbit
is aligned with the star --- the orbit is viewed nearly face-on.
The premise of good alignment was reasonable at the time
but was eventually undermined with the discovery 
of severe misalignments between stars and the orbits of short-period giant planets
\citep[see][for a review]{Albrecht+2022}.

\cite{Kiefer+2019} overcame the impasse
using information from Gaia's first data release.
Although this data release
did not include time-series astrometry, nor the results of two-body fits, it did
report the `astrometric excess noise', a measure of goodness-of-fit
to a model in which the star has no companions. After performing simulations of possible
orbits and the corresponding levels of astrometric excess noise that they would
produce, \cite{Kiefer+2019} concluded that HD\,114762's orbital inclination
is only 4--6$^\circ$ and the secondary mass
is $0.13\pm 0.03~M_\odot$. 
Later, using DR3 data, \cite{GaiaDR3TwoBody2022} and \cite{Holl+2022}
confirmed that HD\,114762 is a face-on binary star.

Although the saga of HD\,114762\,b as a planet candidate has ended,
a joint analysis of all the available data was performed for the sake
of completeness.
For this case, a minor change was
made to the fitting procedure: the parameters
$\omega+\Omega$ and $\omega-\Omega$ were used in the MCMC analysis rather
than $\omega$ and $\Omega$. This is because for nearly face-on binaries,
$\omega+\Omega$ can be measured precisely even though
$\omega$ and $\Omega$ are degenerate.

Given the large quantity of Doppler data spanning more than 30 years,
the parameters
of the spectroscopic orbit are rigidly nailed down.
The Doppler-only analysis gave $P=83.91713\pm 0.00064$~days,
$K=620.1\pm 0.85$~m/s, and $e=0.3442\pm 0.0012$.
These and the other Doppler parameters were found to be consistent
with the Gaia DR3 two-body orbital solution.
The joint fit
does not exhibit any significant tension.
The orbital inclination is $2.8\pm 0.6$~degrees and the secondary mass
is $0.293^{+0.103}_{-0.056}\,M_\odot$.
One would expect a binary star to have a nonzero flux ratio, and indeed, the
flux ratio was found to be $0.052^{+0.070}_{-0.039}$.

Strong covariances exist between the uncertainties of
the secondary mass, the orbital inclination, and the flux ratio,
for the same reason that they were observed for HD\,132406 (Section~\ref{subsec:HD132406}).
To break this degeneracy, a simple mass-luminosity relationship
was employed.
The online version of Table 5 of \cite{PecautMamajek2013} gives the absolute $G$ magnitude
as a function of stellar mass.  For masses between 0.15
and 1.1$\,M_\odot$, the results are well described by a quadratic function,
\begin{equation}
\label{eq:mag}
 M_G = c_0 + c_1 m + c_2 m^2,
\end{equation}
where $m$ is the stellar mass (in solar masses) and the best-fit
coefficients are $c_0$ = 14.14, $c_1= -12.191$, and $c_2=2.713$.

The joint fit was repeated, this time with a Gaussian prior
on the magnitude difference between the two stars
with a mean determined by the application of Eqn.~\ref{eq:mag} to both
stars, and a standard deviation of 0.25 (chosen somewhat arbitrarily).
The prior constraint on the primary star's mass was also loosened
from $1.047\pm 0.040$ to $1.05\pm 0.10$ solar masses to allow
for the possibility that light from the companion affected the
classification of the primary star.
The effect of these changes
was to suppress the solutions with relatively high flux ratios.
The inclination, secondary mass, and flux ratio found through this procedure
were $3.63\pm 0.06$~degrees, $0.215\pm 0.013\,M_\odot$, and $0.17^{+0.14}_{-0.07}\%$,
respectively. The results are given in Table~\ref{tbl:results-hd114762} and depicted
in Figure~\ref{fig:HD114762}, with the corner plot in the Appendix.

\begin{figure*}
\begin{center}
\includegraphics[width=0.9\textwidth]{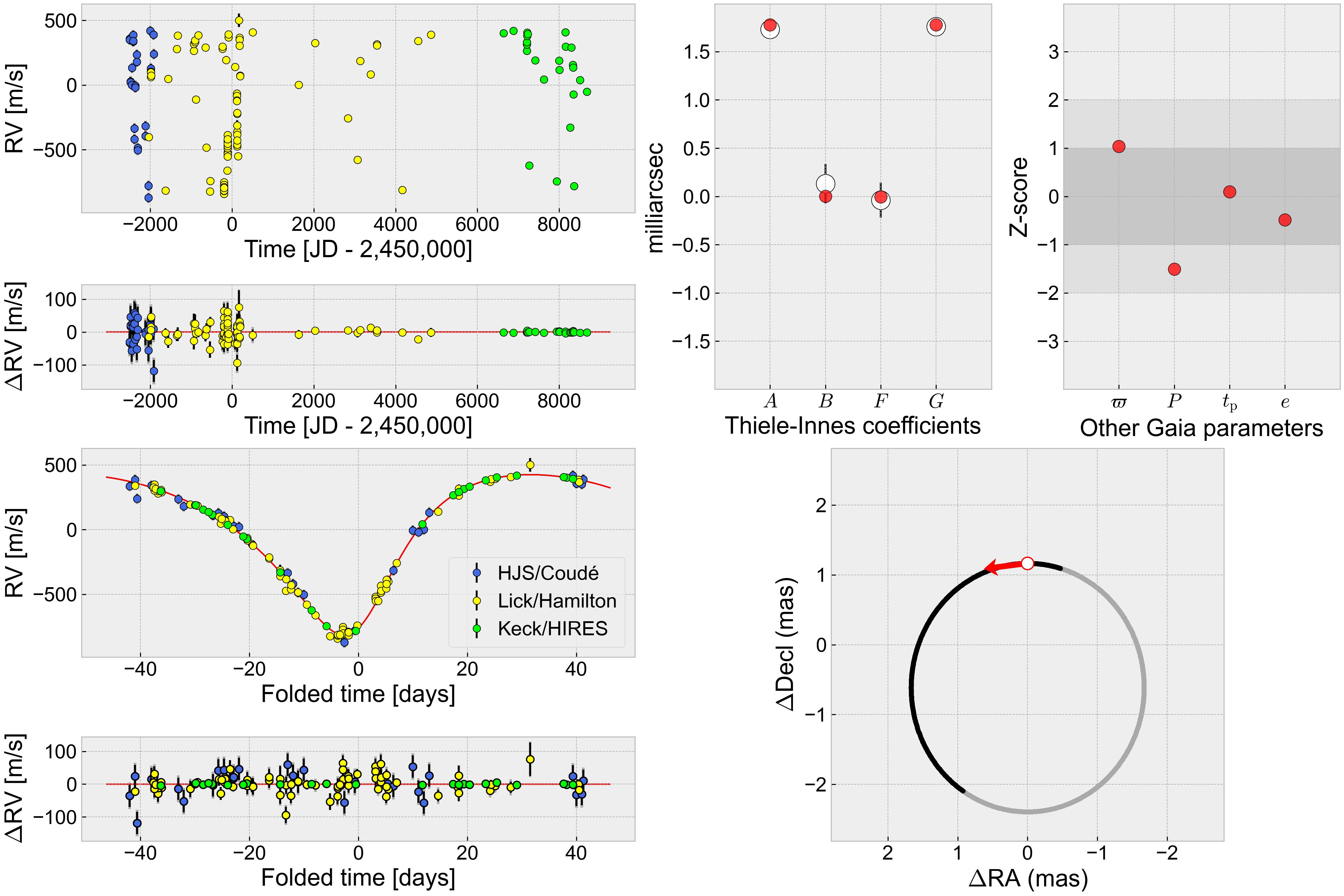}
\end{center}
\caption{\label{fig:HD114762}
{\bf HD\,114762}:
Results of jointly fitting the Doppler and astrometric data. Same format as Figure~\ref{fig:BD-17_0063},
except that the model curve is not shown in the upper left panel because the plotted time range
spans too many cycles for the details to be visible. In the radial-velocity plots, the
blue, yellow, and green points are from the Coude spectrograph on the Harlan
J.\ Smith telescope \citep{Cochran+1991}, the Lick/Hamilton spectrograph \citep{Kane+2011}, and Keck/HIRES \citep{Rosenthal+2021}, respectively.
}
\end{figure*}

In this model, $\cos I = 0.997994\pm 0.000067$, remarkably close to unity.
Only one out of $\approx$500 binaries in a randomly-oriented sample
would be expected to have such a low inclination.
Of course, HD\,114762 was not drawn at random from such a sample.
It drew attention because of the low amplitude
of the Doppler signal, a selection criterion favoring face-on orbits.

\begin{deluxetable*}{lcc}
\label{tbl:results-hd114762}
\tablecaption{Results of jointly fitting the Doppler and Gaia orbital solution for HD\,114762.}
\tablehead{
\colhead{Parameter} & \colhead{Value}            & \colhead{Value} \\[-0.1in]
\colhead{\# of}     & \colhead{(no $M/L$ prior)} & \colhead{(with $M/L$ prior)} 
} 
\startdata
$M$ [$M_\odot$] & $1.046\pm 0.040$                  & $1.00\pm 0.10$  \\
$m$ [$M_\odot$] & $0.293^{+0.103}_{-0.056}$         & $0.215\pm 0.013$ \\
$P$ [days]      & $83.91712\pm 0.00064$             & $83.91712\pm 0.00064$ \\
$t_{\rm p}$ [days] & $-30.798\pm 0.048$             & $-30.795\pm 0.048$ \\
$e$             & $0.3442\pm 0.0012$                & $0.3442\pm 0.0012$ \\
$\cos I$        & $0.99877^{+0.00049}_{-0.00054} $  & $0.997994\pm 0.000067$ \\
$\omega+\Omega$ [rad] & $6.283\pm 0.013$            & $6.283\pm 0.013$ \\
$\omega-\Omega$ [rad] & $0.830\pm 0.014$            & $0.831\pm 0.014$ \\
$\varpi$ [mas]  & $25.35\pm 0.035$                  & $25.35\pm 0.035$ \\
$\varepsilon$ [$\%$] & $5.2^{+7.1}_{-3.9}$          & $0.17^{+0.14}_{-0.07}$
\enddata
%\tablerefs{reference list}
\tablecomments{The $M/L$ prior refers to a prior constraint on the relationship
between the mass ratio and the flux ratio in the Gaia $G$ band. See Eqn.~(\ref{eq:mag}).}
\end{deluxetable*}

\section{Discussion}
\label{sec:discussion}

Exoplanetary systems that can be studied with more than one technique are
especially valuable. The data from different techniques can validate and reinforce each other,
while also providing more powerful constraints on the system's parameters.
At the moment, the most common combination is the pairing of the Doppler and transit techniques.
The Doppler technique supplies $m\sin I$, and the transit
technique ensures $I\approx 90^\circ$ while also giving access to the planet's radius.
According to the NASA Exoplanet Archive, there are 838 `confirmed' planets that
have been detected by both the Doppler and transit techniques.
The next most common combination is that of the Doppler and astrometric techniques.
Until very recently, almost all such systems were Doppler planets
for which astrometric motion was detected
with the Hubble Space Telescope
Fine Guidance Sensors \citep[see, e.g.,][]{Benedict+2002,McArthur+2010},
or through the comparison of Hipparcos and Gaia positions and proper motions
\citep{Li+2021}. Prior to Gaia DR3, there were about 15 objects in this category,
with the exact number depending
on the upper mass limit chosen for planets.
A few directly-imaged planets have also been detected with the Doppler method \citep[see, e.g.,][]{Ruffio+2021}.

Now, astrometric information is available for 73 planets and candidate planets,
of which 9 were already known to exist from Doppler surveys.
Simulations of the Gaia survey suggest that $\sim$10$^4$ planets will eventually
be detectable with Gaia data \citet{Perryman+2014}.
Precise Doppler observations will play an important role in the validation
and characterization of the planets with bright host stars, as emphasized by \citet{GaiaDR3TwoBody2022} and
\citet{Holl+2022}, and as demonstrated in this study.

For now, the publicly available Gaia information is limited to the results
of fitting a two-body model to the data, rather than the time-series
astrometry. When the Gaia orbital parameters have nearly Gaussian uncertainties
and the star's reflex motion 
is dominated by the effect of a single giant planet, there is no obstacle
to combining the Doppler and Gaia information and obtaining good constraints
on the three-dimensional orbit as well as the planet's mass, as was the case
for BD\=/17\,0063, HD\,81040, and HD\,132406. In some cases,
though, the interpretation of the data will be more difficult
because of non-Gaussian uncertainties,
as was the case for HD\,175167 and HR\,810,
or the presence of multiple planetary signals, as is the case
for HD\,111232 and possibly for HIP\,66074.

Progress is possible now, but much will have to wait until Gaia DR4 when the
time-series astrometric data will become available.
In the meantime, it would be useful to conduct long-term Doppler monitoring of
stars known to have giant planets that are potentially detectable with Gaia.
Too many giant planets have been discovered with the Doppler method and then
ignored for a decade or more.
Obtaining at least a little more Doppler data over timescales of a few years
before DR4 would enhance our ability to interpret the Gaia data and make the most
of the enormous exoplanet potential of the Gaia mission.

\begin{acknowledgements}
This work would not have been possible
without the hard work over many years of the Gaia team,
who have delivered a dataset with a breathtaking
scope of applications.
The author is also grateful to the anonymous referee for a timely and helpful
report, and to D.\ Foreman-Mackey for his development of the {\tt emcee}
and {\tt corner} Python codes.
This work has made use of data from the European Space Agency (ESA) mission
Gaia (\url{https://www.cosmos.esa.int/gaia}), processed by the Gaia
Data Processing and Analysis Consortium (DPAC,
\url{https://www.cosmos.esa.int/web/gaia/dpac/consortium}). Funding for the DPAC
has been provided by national institutions, in particular the institutions
participating in the Gaia Multilateral Agreement.
This research also made use of the \cite{NEA}, which is operated by the California Institute of Technology, under contract with the National Aeronautics and Space Administration under the Exoplanet Exploration Program.
\end{acknowledgements}

\appendix

\section{Corner plots}

\begin{figure*}
\begin{center}
\includegraphics[width=0.9\textwidth]{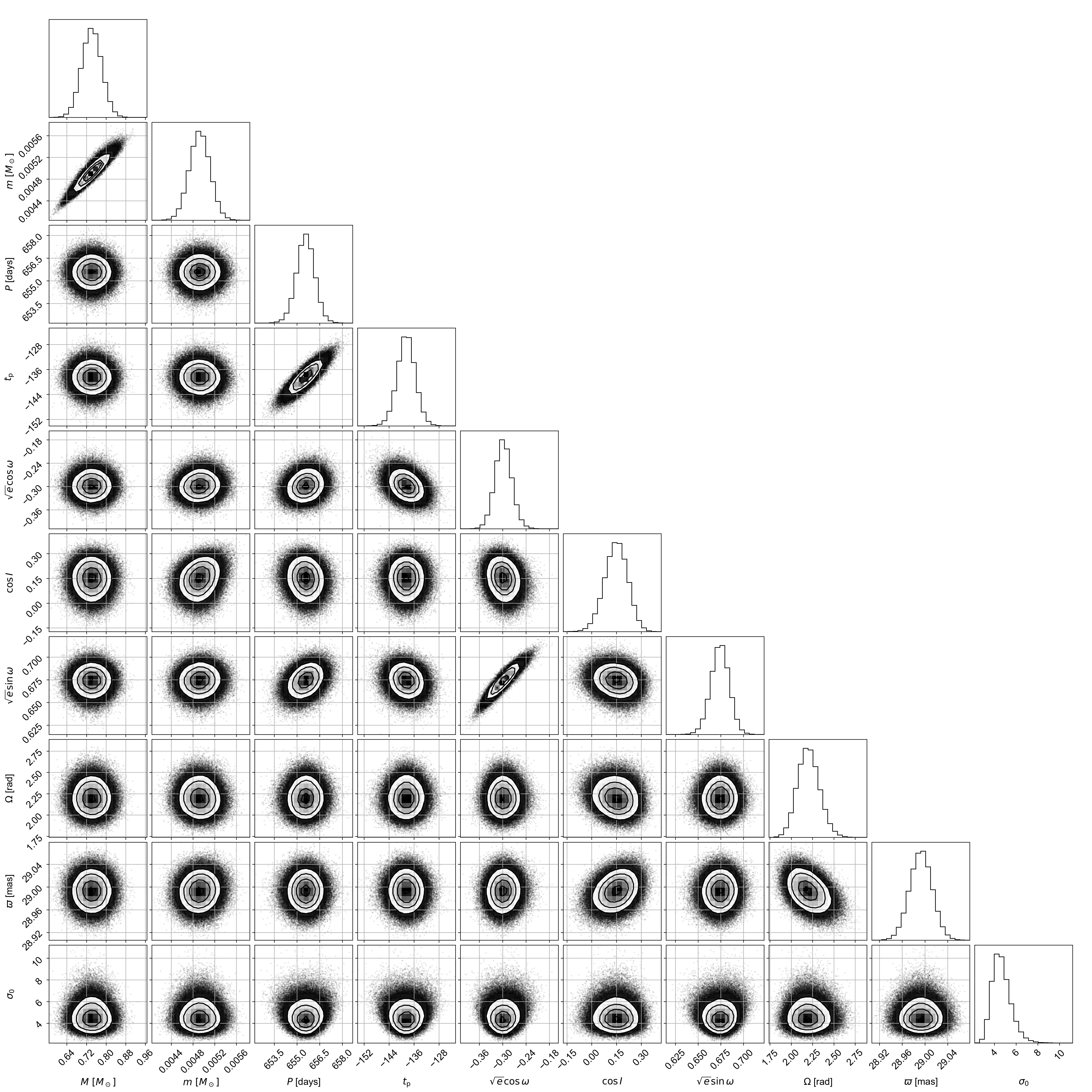}
\end{center}
\caption{\label{fig:BD-17_0063-corner}
{\bf BD\=/17\,0063}}
\end{figure*}

\begin{figure*}
\begin{center}
\includegraphics[width=0.9\textwidth]{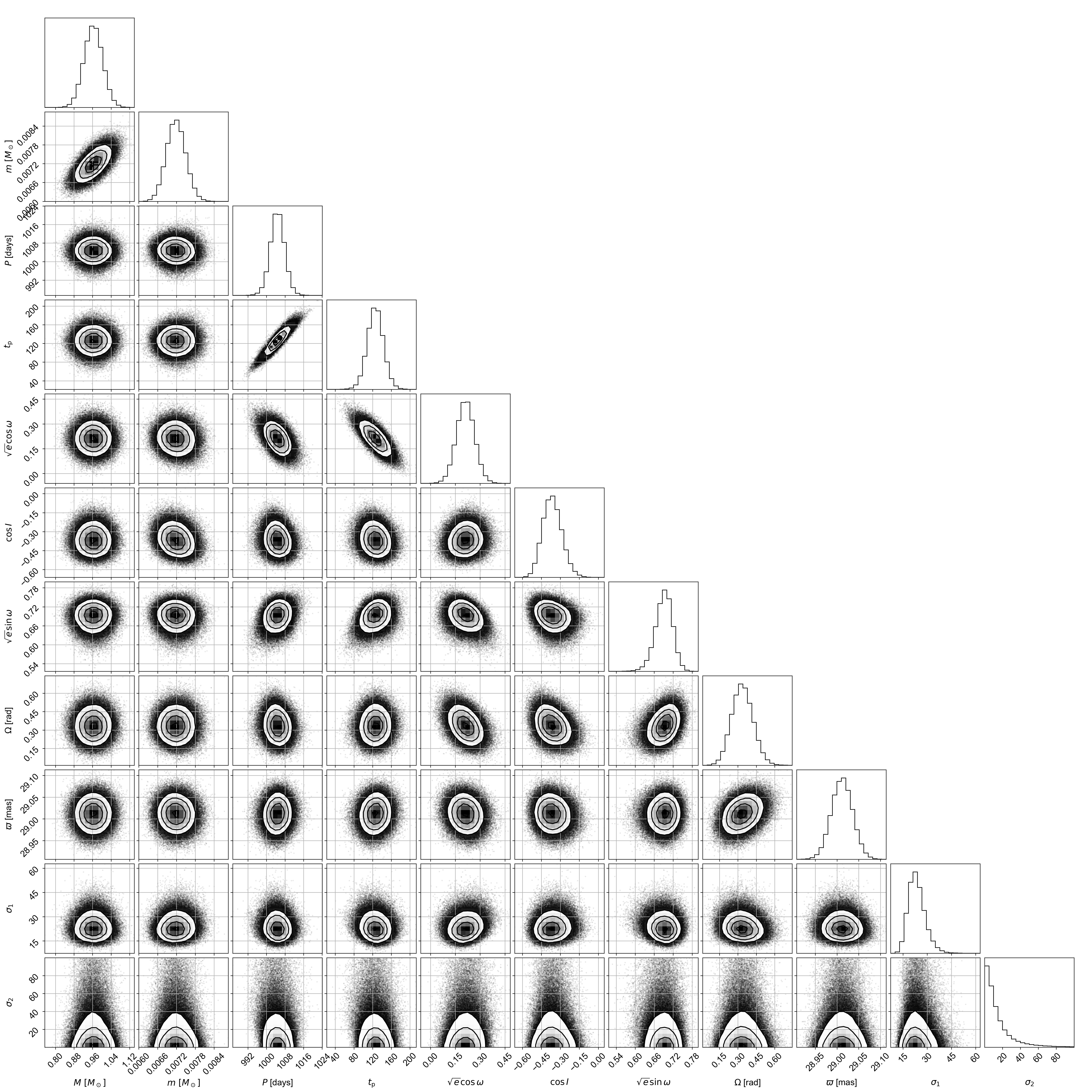}
\end{center}
\caption{\label{fig:HD81040-corner}
{\bf HD~81040}}
\end{figure*}

\begin{figure*}
\begin{center}
\includegraphics[width=0.9\textwidth]{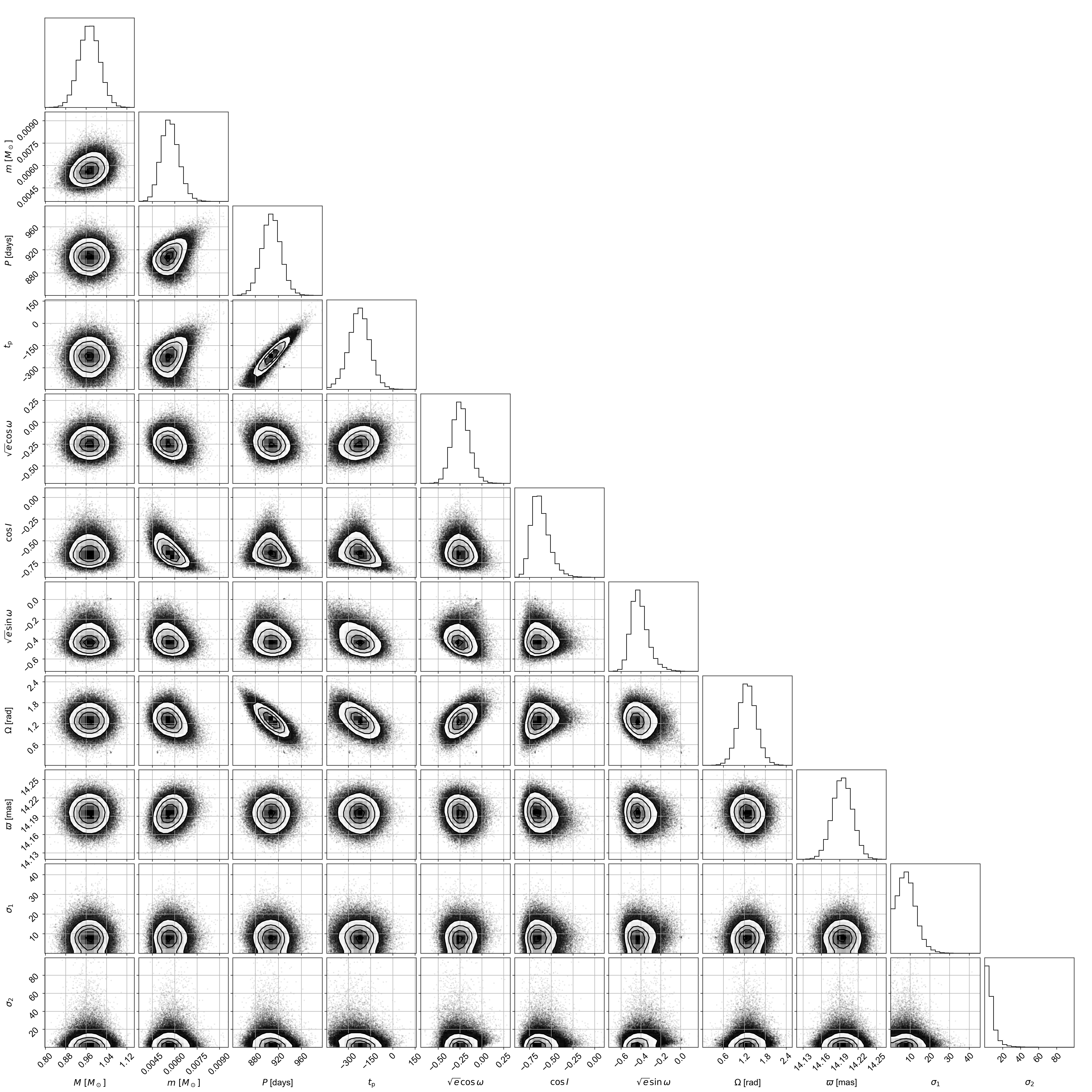}
\end{center}
\caption{\label{fig:HD132406-corner}
{\bf HD~132406}}
\end{figure*}

\begin{figure*}
\begin{center}
\includegraphics[width=0.9\textwidth]{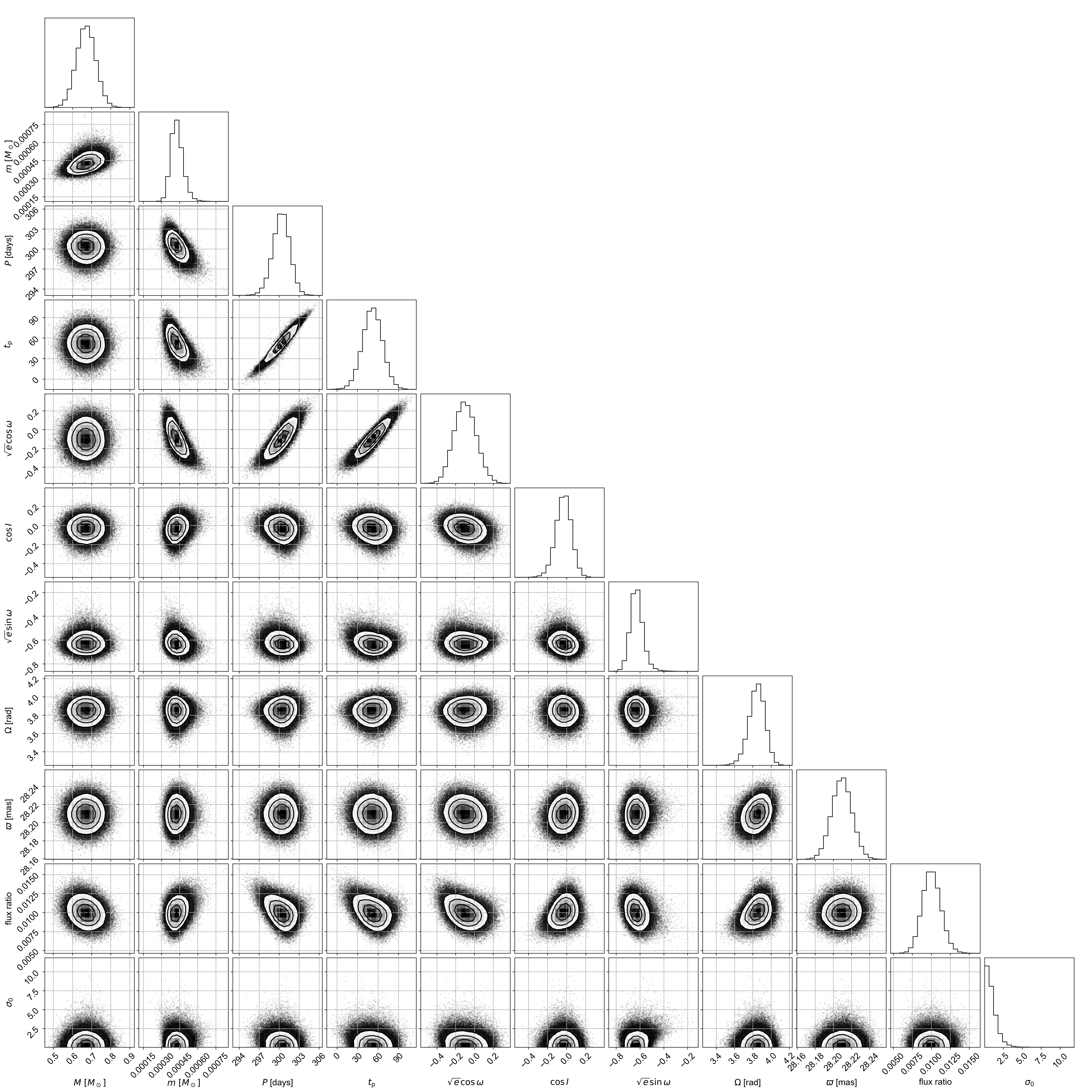}
\end{center}
\caption{\label{fig:HIP66074-corner}
{\bf HIP\,66074}}
\end{figure*}

\begin{figure*}
\begin{center}
\includegraphics[width=0.9\textwidth]{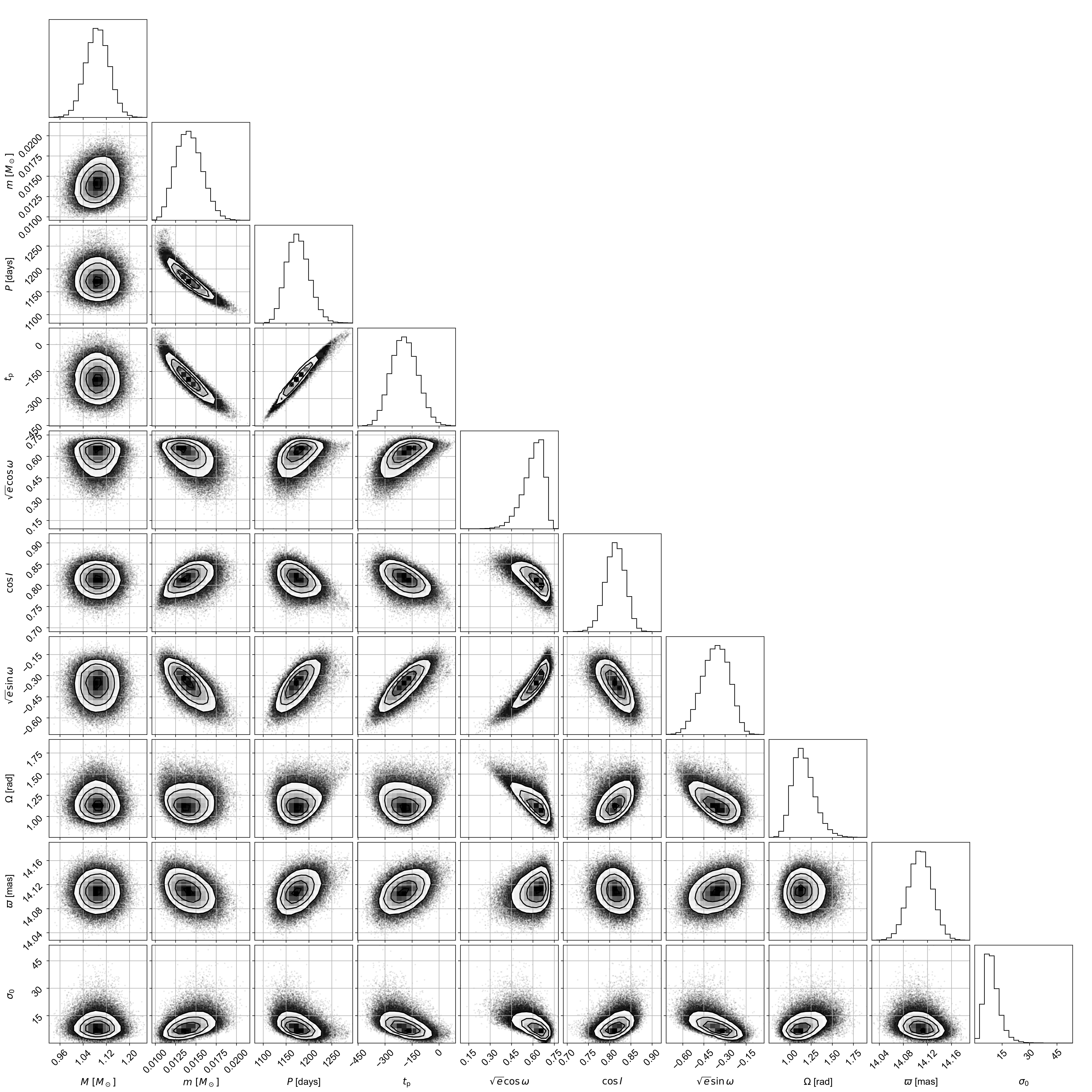}
\end{center}
\caption{\label{fig:HD175167-corner}
{\bf HD\,175167}}
\end{figure*}

\begin{figure*}
\begin{center}
\includegraphics[width=0.9\textwidth]{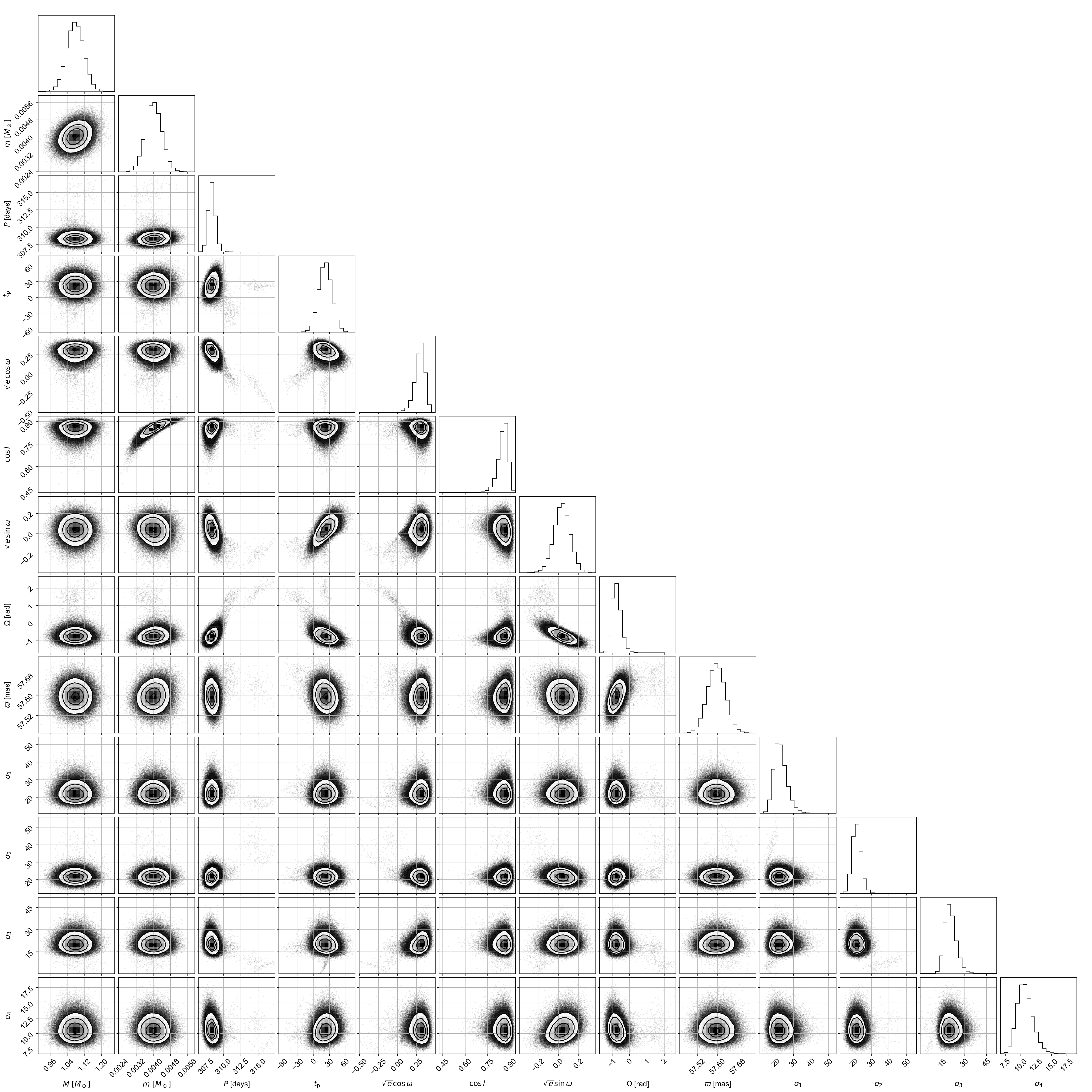}
\end{center}
\caption{\label{fig:HR810-corner}
{\bf HR\,810}}
\end{figure*}

\begin{figure*}
\begin{center}
\includegraphics[width=0.9\textwidth]{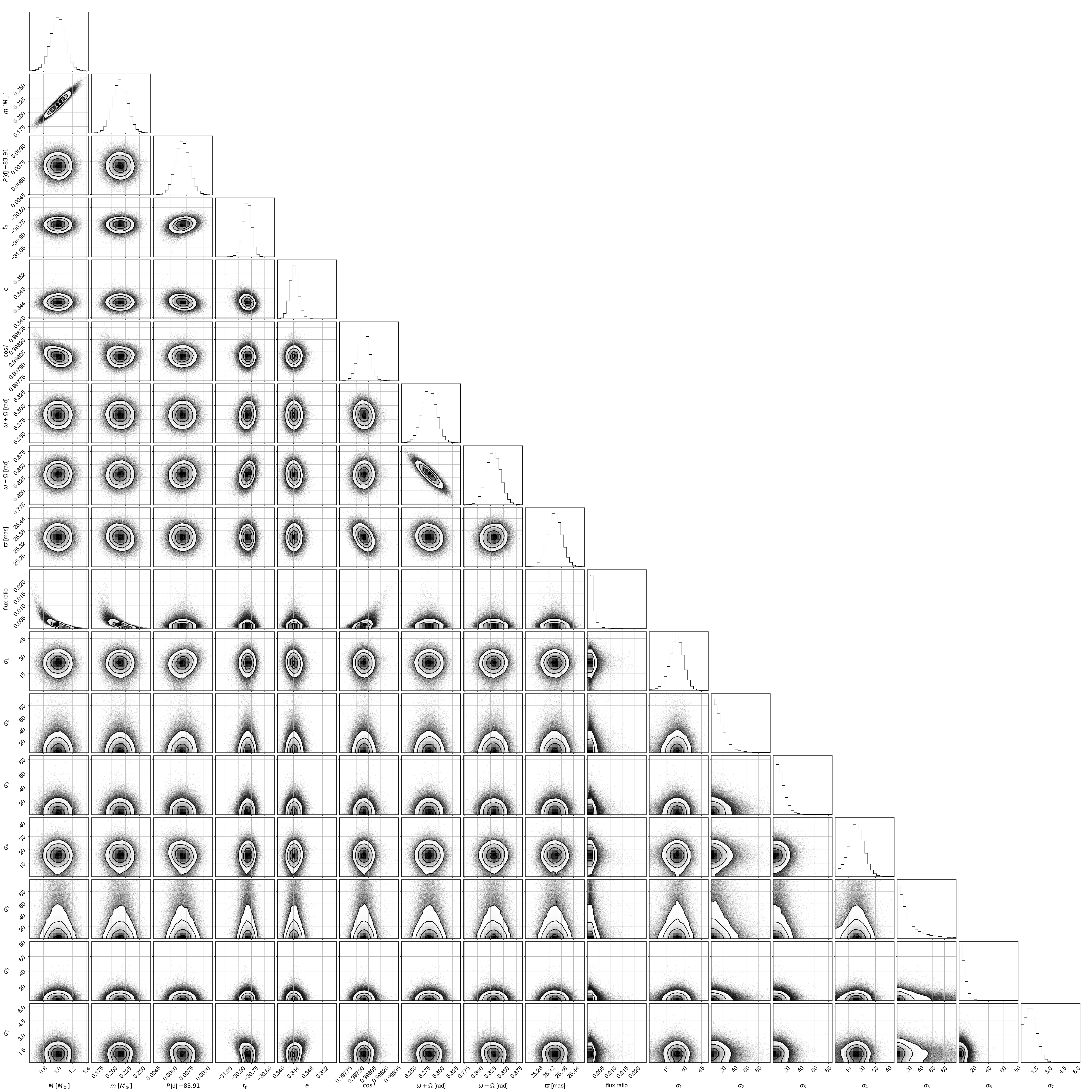}
\end{center}
\caption{\label{fig:HD114762-corner}
{\bf HD\,114762}}
\end{figure*}

Contained here are plots of the joint {\it a posteriori} probability distribution
of the parameters for each system, based on the simultaneous fit to the Doppler data
and Gaia orbital solution, and depicted as a `corner plot' of 2-d distributions.
The results for BD\=/17\,0063, HD~81040, HD~132406, HIP\,66074, HD\,175167, HR\,810, and HD\,114762
are shown in Figures~\ref{fig:BD-17_0063-corner}, \ref{fig:HD81040-corner}, \ref{fig:HD132406-corner},
\ref{fig:HIP66074-corner}, \ref{fig:HD175167-corner}, \ref{fig:HR810-corner}, and \ref{fig:HD114762-corner}, respectively.

\bibliography{ms}{}
\bibliographystyle{aasjournal}

\end{document}